\documentclass[manuscript]{aastex}
\usepackage{ulem}
\usepackage{color}

\def\n2h{N$_2$H$^+$ }

\def\deg{$^\circ$}

\def\13co{$^{13}$CO}
\def\c18o{C$^{18}$O}

\def\obj{$\rho$ Ophiuchi molecular cloud complex }

\shorttitle{The $\rho$ Oph chemical Characteristics}
\shortauthors{Pan, Z., Li, D., Chang, Q., Qian, L., Bergin, E. A., Wang, J.}

\begin{document}

\title{Large-Scale Spectroscopic Mapping of the $\rho$ Ophiuchi Molecular Cloud Complex I. The C$_{2}$H to N$_2$H$^+$ Ratio as a Signpost of Cloud Characteristics}

\author{Zhichen Pan\altaffilmark{1}, Di Li\altaffilmark{1,2,3}, Qiang Chang\altaffilmark{4}, Lei Qian\altaffilmark{1,2}, Edwin A. Bergin\altaffilmark{5}, Junzhi Wang\altaffilmark{6}}
\altaffiltext{1}{National Astronomical Observatories, Chinese Academy of Sciences, A20 Datun Road, Chaoyang District, Beijing 100012, China. Email: panzc@nao.cas.cn, dili@nao.cas.cn}
\altaffiltext{2}{Key Laboratory of Radio Astronomy, Chinese Academy of Sciences, Nanjing 210008, China}
\altaffiltext{3}{University of Chinese Academy of Sciences 19 A Yuquan Rd, Shijingshan District, Beijing, P.R.China 100049}
\altaffiltext{4}{Xinjiang Astronomical Observatory, Chinese Academy of Sciences, 150 Science 1-Street, Urumqi 830011, China}
\altaffiltext{5}{Department of Astronomy, University of Michigan, 500 Church St., Ann Arbor, MI 48109, USA}
\altaffiltext{6}{Shanghai Astronomical Observatory, Chinese Academy of Sciences, 80 Nandan Road, Shanghai 200030, China}

\begin{abstract}
    We present 2.5-square-degree C$_{2}$H N=1-0 and N$_2$H$^+$ J=1-0 maps of the $\rho$ Ophiuchi molecular cloud complex.
    These are the first large-scale maps of the $\rho$ Ophiuchi molecular cloud complex with these two tracers.
    The C$_{2}$H emission is spatially more extended than the N$_2$H$^+$ emission.
    One faint N$_2$H$^+$ clump Oph-M and one C$_{2}$H ring Oph-RingSW are identified for the first time.
    The observed C$_{2}$H to N$_{2}$H$^{+}$ abundance ratio ([C$_{2}$H]/[N$_{2}$H$^{+}$]) varies between 5 and 110.
    We modeled the C$_{2}$H and N$_2$H$^+$ abundances with 1-D chemical models
    which show a clear decline of [C$_2$H]/[N$_2$H$^+$] with chemical age.
    Such an evolutionary trend is little affected by temperatures when they are below 40 K.
    At high density (n$_H$ $>$ 10$^5$ cm$^{-3}$),
    however, the time it takes for the abundance ratio to drop at least one order of magnitude becomes less than the dynamical time (e.g., turbulence crossing time $\rm \sim$10$^5$ years).
    The observed [C$_2$H]/[N$_2$H$^+$] difference between L1688 and L1689 can be explained by L1688 having chemically younger gas in relatively less dense regions.
    The observed [C$_{2}$H]/[N$_{2}$H$^{+}$] values are the results of time evolution, accelerated at higher densities.
    For the relative low density regions in L1688 where only C$_2$H emission was detected, the gas should be chemically younger.

\end{abstract}


\section{INTRODUCTION}

  With a distance of 120.0$^{+4.5}_{-4.2}\ \  \rm pc$ (Loinard et al.\ 2008),
  the \obj (hereafter $\rho$-Oph MCC), also called the Ophiuchus Molecular Cloud (e.g.,\ Pattle et al.\ 2015),
  is one of the most active star-forming regions within several hundred parsecs of the sun (Lada \& Lada 2003).
  At such a close distance, the 60$''$ beam size (at $\sim \rm$100 GHz) of a ten-meter class millimeter radio telescope corresponds to 0.035 pc.
  The $\rho$-Oph MCC consists of Lynds 1688 (hereafter L1688), L1689, and a number of filamentary clouds (L1709, L1740, L1744, L1755, and L1765).
  These filamentary clouds extending from L1688, are known as the streamers or the cobwebs of the $\rho$-Oph MCC (Nutter et al.\ 2006).
  L1688 is an intermediate mass star-forming region,
  filling the gap between low-mass star-forming regions, such as Taurus, and more massive star-forming regions, exemplified by the Orion molecular cloud (Padgett et al.\ 2008).
  Motte et al.\ (1998) identified 13 dense condensations in L1688 from their 1.3 mm dust continuum map with an effective beam resolution of 13$''$ (FWHM and all instances hereafter).
  Johnstone et al.\ (2004) identified two more objects outside the boundary of previous maps in one 850 $\mu$m map with a 40 mJy beam$^{-1}$ sensitivity and a 14$''$ beam size.
  Stanke et al.\ (2006) presented a 1.2 mm dust continuum L1688 map with a $\rm \sim$24$''$ resolution and a noise level of the order of 10 mJy beam$^{-1}$.
  They cleaned their map down to the noise level by using wavelet analysis and identified 143 sources.
  A possible explanation as to why the number of sources Stanke et al.\ (2006) identified are much larger than those identified by Motte et al.\ (1998) is that
  the condensations in Motte et al.\ (1998) are split into multiple sources.
  For example,
  Oph-B1 was identified as a dense condensation with four starless clumps (B1-MMS1, B1-MMS2, B1-MSS3, and B1-MMS4) in Motte et al.\ (1998)
  while at least six sources (MMS20, MMS23, MMS27, MMS28, MMS66, MMS86, and maybe MMS49) were identified by Stanke et al.\ (2006).
  Other 1.2 mm dust sources in Stanke et al.\ (2006) are isolated sources or faint and extended dust emission regions.
  L1689 is to the east of L1688, associated with less star formation (Loren et al.\ 1990).
  To further examine the similarities and differences between L1688 and L1689, we explore here the utilities of C$_{2}$H and N$_{2}$H$^{+}$ as tracers.

  C$_{2}$H is a tracer of relatively high density gas with a critical density of $1-4 \times 10^{5} $ cm$^{-3}$ (see Appendix A for details).
  It can also be used as an efficient tracer of extended gas due to its lower effective excitation density (cf.\ Shirley 2015).
  The abundances of C$_{2}$H decrease from diffuse gas to dense gas making it a potential probe of chemical evolution.
  In diffuse gas, CO is dissociated by UV photons, releasing carbon for C$_2$H formation and thus sustaining a higher C$_2$H abundance (Li et al.\ 2012).
  In dense gas, without CO dissociation, the C$_{2}$H abundance decreases.
  In contrast, N$_2$H$^+$ is a well-known dense gas tracer with a critical density of $0.5-1.5 \times 10^{6} $ cm$^{-3}$ (Appendix A).

  In this paper,
  we present the first large-scale (2.5 square degree) C$_{2}$H N=1-0 and N$_2$H$^+$ J=1-0 maps,
  which cover most high extinction regions with $A_V$ $>$ $10^{\rm m}$ in L1688 and L1689,
  including clumps in earlier studies (e.g.,\ Di Francesco et al.\ 2004, Andr\'{e} et al.\ 2007). 

  In Section 2, we present the C$_{2}$H and N$_2$H$^+$ observations,
  data reduction,
  and the Infrared Astronomy Satellite (IRAS) dust temperature map.
  The calculation of the C$_{2}$H to N$_{2}$H$^{+}$ abundance ratio ([C$_{2}$H]/[N$_{2}$H$^{+}$]) is presented in Section 3. 
  Our chemical model and interpretation are given in Section 4.
  Comparative studies of C$_{2}$H and N$_{2}$H$^{+}$ in terms of centroid velocity, line width, and [C$_{2}$H]/[N$_{2}$H$^{+}$] are in Section 5.
  Conclusions are presented in Section 6.
  Details of critical density estimation and column density calculation are given in Appendices.

\section[]{OBSERVATION AND DATA REDUCTION}

\subsection[]{The C$_{2}$H and N$_2$H$^+$ Data}

  The $\rho$-Oph MCC was mapped with the Delingha 13.7 m telescope of Qinghai Station\footnote{http://www.dlh.pmo.ac.cn},
  Purple Mountain Observatory (PMO),
  Chinese Academy of Sciences (CAS).
  The observations were carried out in position switching mode in December of 2012 and January, March, and April of 2013.

  At Delingha, the $\rho$-Oph MCC stays above 20$^{\circ}$ elevation with a maximum of 28$^{\circ}$ for approximately 5 hours per day.
  To obtain low system temperatures,
  we only took data for 2 hours around maximum source elevation.

  The 3$\times$3 Superconductor-Insulator-Superconductor (SIS) receiver array of the Delingha telescope (Shan et al.\ 2012; Zuo et al.\ 2011) can observe 85-115 GHz.
  The backend consists of 18 FFT spectrometers, each with 16384 channels.
  The backend operates in two modes, a 200 MHz bandwidth with 12.2 kHz channel width and a 1000 MHz bandwidth with 61.0 kHz channel width.

  As a first test observation, we performed OTF (On The Fly) mapping of the Oph-A region with a 200 MHz bandwidth ($\Delta$ v $\rm \sim$ 0.05 km s$^{-1}$)
  to see if the velocity resolution in the 1000 MHz bandwidth ($\Delta$ v $\rm \sim$0.20 km s$^{-1}$) is suitable for fitting the centroid velocity and line width.
  The test observation covered a $15'\times 15'$ square region centered on the position of Oph-A with the highest 2MASS extinction.
  Hyperfine structure fitting was done.
  The data were then smoothed to a 0.20 km s$^{-1}$ velocity resolution before another round of fitting was performed.
  A comparison between these two fitting results showed that the hyperfine structure fitting of spectra with strong emission is affected little by this smoothing
  and thus the 0.20 km s$^{-1}$ velocity resolution is sufficient for observing N$_2$H$^+$ in the $\rho$-Oph MCC and analysis using hyperfine fitting.
  For example, the original spectrum with a $\rm \sim$4 K T$_{A}^{*}$ peak at the position 16$^{\rm h}$26$^{\rm m}$26.5$^{\rm s}$, -24\deg24$'$09$''$ (J2000) showed
  a centroid velocity 3.60$\pm$0.01 km s$^{-1}$,
  a line width 0.60$\pm$0.03 km s$^{-1}$,
  an optical depth 0.73$\pm$0.2,
  and an excitation temperature 5.7$^{+3.4}_{-1.9}$ K from hyperfine fitting.
  After smoothing to 0.2 km s$^{-1}$,
  the centroid velocity was 3.60$\pm$0.01 km s$^{-1}$,
  the line width was 0.60$\pm$0.03 km s$^{-1}$,
  the optical depth was 0.73$\pm$0.25,
  and the excitation temperature was 5.6$^{+4.4}_{-2.2}$ K.
  In this example, only the uncertainties became slightly larger.
  Compared with the crowded N$_2$H$^+$ hyperfine structures, the C$_{2}$H spectra have six well-separated hyperfine components.
  Observing these six components requires a much lower velocity resolution in a wider bandwidth.
  The 1000 MHz mode provides enough velocity resolution for C$_{2}$H hyperfine structure fitting.
  For spectra with weak emission, however, we performed hyperfine fitting and found that the results were affected by the smoothing.
  In the following,
  we only applied the resulting opacity correction to N$_2$H$^+$ spectra with an integrated intensity of 3 K km s$^{-1}$ or higher and to C$_{2}$H spectra above 2 K km s$^{-1}$.
  Low opacities were assumed for spectra with weak emission.

  The COordinated Molecular Probe Line Extinction Thermal Emission Survey of Star Forming Regions (Ridge et al.\ 2006, hereafter COMPLETE)
  provides the 2MASS NICER extinction map and IRAS dust temperature map of the $\rho$-Oph MCC.
  These maps were used for observation region selection.
  In L1688, six $30'\times 30'$ areas were selected,
  covering most regions with extinction A$_{V}>$10$^{\rm m}$.
  L1689 was covered by four $30'\times 30'$ areas next to the observation region of L1688.
  Reference positions were chosen from the COMPLETE $\rho$-Oph MCC $^{12}$CO and \13co maps.
  Tables \ref{coordinate} and \ref{freq_set} show the details of observation.

\begin{table}
\centering
\begin{tiny}
  \caption{Observation information of ten 30$'\times$30$'$ OTF scans.}
\vspace{1em}
  \begin{tabular}{lcccccccc}
  \hline
  No.    & Center R.A.                         & Center Decl.          &  Ref. Position                    & Ref. Position     & Scan Rate      & Sample Interval & Scan Time   & Total Scan Time \\
         & (J2000)                           & (J2000)             &  R.A. (J2000)                       & Decl. (J2000)       &(arcsec s$^{-1}$) & (s)             & (minutes)      & (minutes $\times$ repeats)          \\
  \hline
  01     & 16$^{\rm h}$26$^{\rm m}$16$^{\rm s}$  & -24\deg20$'$54$''$  &  16$^{\rm h}$29$^{\rm m}$29$^{\rm s}$  & -24\deg29$'$31$''$     & 50         & 0.3             & 85             & 85 minute $\times$ 4                   \\
  02     & 16$^{\rm h}$28$^{\rm m}$16$^{\rm s}$  & -24\deg20$'$54$''$  &  16$^{\rm h}$30$^{\rm m}$32$^{\rm s}$  & -24\deg07$'$28$''$     & 50         & 0.3             & 85             & 85 minute $\times$ 4                   \\
  03     & 16$^{\rm h}$24$^{\rm m}$16$^{\rm s}$  & -24\deg20$'$54$''$  &  16$^{\rm h}$26$^{\rm m}$18$^{\rm s}$  & -23\deg46$'$10$''$     & 50         & 0.3             & 85             & 85 minute $\times$ 4                   \\
  04     & 16$^{\rm h}$26$^{\rm m}$16$^{\rm s}$  & -24\deg50$'$54$''$  &  16$^{\rm h}$30$^{\rm m}$32$^{\rm s}$  & -24\deg07$'$28$''$     & 50         & 0.3             & 85             & 85 minute $\times$ 4                   \\
  05     & 16$^{\rm h}$28$^{\rm m}$16$^{\rm s}$  & -24\deg50$'$54$''$  &  16$^{\rm h}$30$^{\rm m}$32$^{\rm s}$  & -24\deg07$'$28$''$     & 75         & 0.2             & 45             & 60 minute $\times$ 6                   \\
  06     & 16$^{\rm h}$24$^{\rm m}$16$^{\rm s}$  & -24\deg50$'$54$''$  &  16$^{\rm h}$26$^{\rm m}$18$^{\rm s}$  & -23\deg46$'$10$''$     & 75         & 0.2             & 45             & 60 minute $\times$ 6                   \\
  07     & 16$^{\rm h}$30$^{\rm m}$16$^{\rm s}$  & -24\deg30$'$00$''$  &  16$^{\rm h}$30$^{\rm m}$16$^{\rm s}$  & -24\deg18$'$00$''$     & 75         & 0.2             & 45             & 60 minute $\times$ 6                   \\
  08     & 16$^{\rm h}$32$^{\rm m}$16$^{\rm s}$  & -24\deg30$'$00$''$  &  16$^{\rm h}$32$^{\rm m}$16$^{\rm s}$  & -24\deg18$'$00$''$     & 75         & 0.2             & 45             & 60 minute $\times$ 6                   \\
  09     & 16$^{\rm h}$30$^{\rm m}$16$^{\rm s}$  & -25\deg00$'$00$''$  &  16$^{\rm h}$30$^{\rm m}$16$^{\rm s}$  & -24\deg18$'$00$''$     & 75         & 0.2             & 45             & 60 minute $\times$ 6                   \\
  10     & 16$^{\rm h}$32$^{\rm m}$16$^{\rm s}$  & -25\deg00$'$00$''$  &  16$^{\rm h}$32$^{\rm m}$16$^{\rm s}$  & -24\deg18$'$00$''$     & 75         & 0.2             & 45             & 60 minute $\times$ 6                   \\

\hline
  \label{coordinate}
\end{tabular}
\end{tiny}
\end{table}

\begin{table}
\centering
\begin{tiny}
  \caption{Frequency settings of the observation.
  The frequencies and relative intensities of C$_{2}$H are from Tucker et al.\ (1974),
  the frequencies and relative intensities of N$_2$H$^+$ are from Keto \& Rybicki (2010).}
\vspace{1em}
  \begin{tabular}{lcccccc}
  \hline
  Lines                                            & Rest Frequency       & Relative Intensity  \\
                                                   &     (MHz)            &         \\
  \hline
  C$_{2}$H N=1-0 J=$\frac{3}{2}$-$\frac{1}{2}$ F=1-1   & 87284.38            &   4.25                   \\
  C$_{2}$H N=1-0 J=$\frac{3}{2}$-$\frac{1}{2}$ F=2-1   & 87317.05            &   41.67                   \\
  C$_{2}$H N=1-0 J=$\frac{3}{2}$-$\frac{1}{2}$ F=1-0   & 87328.70            &   20.75                   \\
  C$_{2}$H N=1-0 J=$\frac{1}{2}$-$\frac{1}{2}$ F=1-1   & 87402.10            &   20.75                   \\
  C$_{2}$H N=1-0 J=$\frac{1}{2}$-$\frac{1}{2}$ F=0-1   & 87407.23            &   8.33                   \\
  C$_{2}$H N=1-0 J=$\frac{1}{2}$-$\frac{1}{2}$ F=1-0   & 87446.42            &   4.25                   \\
  N$_2$H$^+$ J=1-0 F$_{1}$=1-1 F=0-1                     & 93171.6086            &   0.33334                  \\
  N$_2$H$^+$ J=1-0 F$_{1}$=1-1 F=2-1                     & 93171.9054            &   1.66667                  \\
  N$_2$H$^+$ J=1-0 F$_{1}$=1-1 F=1-1                     & 93172.0403            &   1.00000                  \\
  N$_2$H$^+$ J=1-0 F$_{1}$=2-1 F=2-1                     & 93173.4675            &   1.66667                   \\
  N$_2$H$^+$ J=1-0 F$_{1}$=2-1 F=3-1                     & 93173.7643            &   2.33333                  \\
  N$_2$H$^+$ J=1-0 F$_{1}$=2-1 F=1-1                     & 93173.9546            &   1.00000                   \\
  N$_2$H$^+$ J=1-0 F$_{1}$=0-1 F=1-1                     & 93176.2527            &   1.00000                  \\
\hline
\label{freq_set}
\end{tabular}
\tablenotetext{}{Note: C$_{2}$H is in lower sideband and N$_2$H$^+$ is in upper sideband.}
\end{tiny}
\end{table}

  The original OTF data were regridded with a pixel size of 30$'' \times$ 30$''$.
  In the regridding procedure, spectra with extremely bad baselines or high Root Mean Square (RMS) values were removed, since these spectra may be caused by hardware errors.
  A 60\% main beam efficiency was used in the calibration.
  The data were uploaded to the Millimeter Wave Radio Astronomy Database\footnote{http://www.radioast.csdb.cn}$^{\rm ,}$\footnote{http://www.dlh.pmo.cas.cn}$^{\rm ,}$\footnote{http://www.csdb.cn}.
  The astronomical software package Gildas/CLASS\footnote{http://www.iram.fr/IRAMFR/GILDAS/} was used for further reduction.

  Given the relatively low elevation of the observations, the spectral baselines are complex.
  A 5$^{th}$ order polynomial was subtracted from each C$_{2}$H spectrum.
  For N$_2$H$^+$ data, a 3$^{rd}$ order polynomial was used.
  The OTF data were combined into one file for regridding.
  In regridding, the velocity resolutions are 0.21 km s$^{-1}$ for C$_{2}$H spectra and 0.20 km s$^{-1}$ for N$_2$H$^+$ spectra.
  After regridding, only spectra with RMS values less than 0.5 K (T$_{A}^{*}$) were selected.
  For C$_{2}$H spectra, velocity ranges of [-50, -40], [-30, 0], and [10, 50] km s$^{-1}$ were used for RMS calculation.
  For N$_2$H$^+$ spectra, the velocity ranges used were [-50, -8] and [15, 50] km s$^{-1}$.
  Though the RMS values of the spectra near map edges were still less than 0.5 K (T$_{A}^{*}$), the baselines of these spectra were too complex for polynomial fitting.
  Hence, spectra near map edges were rejected.

  The integrated intensity maps of C$_{2}$H and N$_2$H$^+$ are shown in Figure \ref{intg}. 
  For C$_2$H, to present a uniform integrated intensity map, we selected two strongest hyperfine components that were widely detected.
  The selected velocity ranges for these two components in spectra were
  [-39.1, -35.1] and [1.9, 5.0]\footnote{N=1-0 J=3$/$2-1$/$2 F=2-1 and N=1-0 J-1$/$2-1$/$2 F=1-1 }km s$^{-1}$.
  {When calculating the N$_2$H$^+$ integrated intensity, we selected three hyperfine components, corresponding to a velocity range of
  [2.0, 6.0]\footnote{J=1-0 F$_1$=2-1 F=2-1, J=1-0 F$_1$=2-1 F=3-1, and J=1-0 F$_1$=2-1 F=1-1, which are three hyperfine components in the main group. } km s$^{-1}$.
  The RMS values of C$_{2}$H and N$_2$H$^+$ integrated intensity maps are 0.8 K km s$^{-1}$ and 0.4 K km s$^{-1}$, respectively.
  A 5$\times$5 pixel boxcar smoothing was then performed to the integrated intensity maps for drawing contours.

  The weather and the relatively low elevation are the main reasons for the variation of RMS values.
  When the observation region is close to its maximum elevation and the weather is good, a typical system temperature is between 160 K and 170 K,
  which increases to 200 K or higher under bad weather or low elevation.
  The variation of the system temperature caused an approximately $\pm$0.1 K km s$^{-1}$ variation of the RMS values of the smoothed C$_{2}$H and N$_2$H$^+$ spectra in different regions.
  The actual sky coverage of each 30$' \times$ 30$'$ OTF map is a little larger.
  Around the edges with overlapped scans, we averaged both data sets to lower the RMS values.

\begin{figure}
\centering
\includegraphics[width=140mm,angle=0]{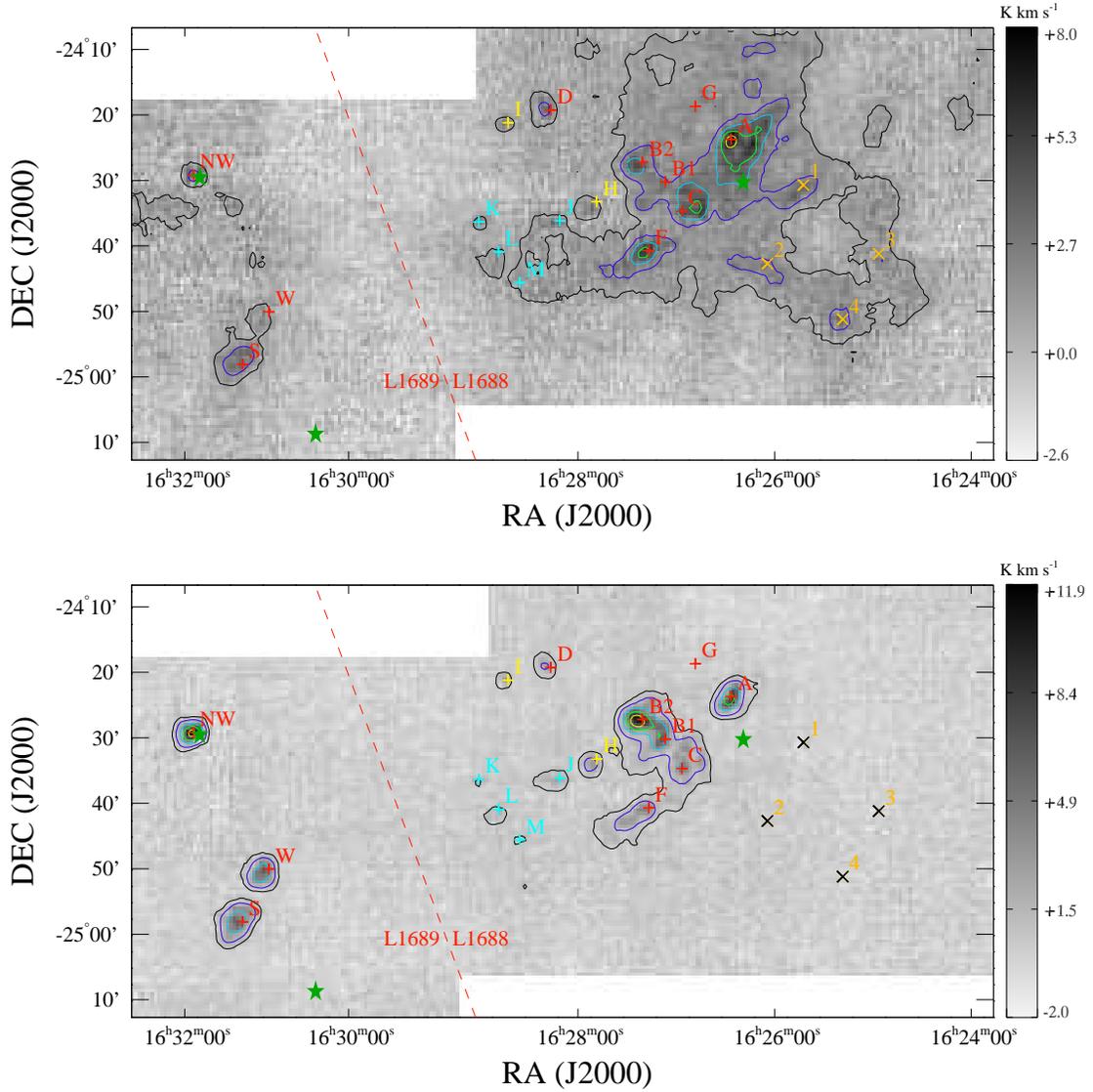}
\vspace{2em}
\caption{\tiny
  C$_{2}$H integrated intensity map (top) and N$_2$H$^+$ integrated intensity map (bottom).
  The data are smoothed with a boxcar average of a 5 $\times$ 5 pixel box only for drawing contours.
  In the C$_{2}$H map, black, purple, blue, green, and yellow contours show the smoothed C$_{2}$H integrated intensities at levels of 1, 2, 3, 4, and 5 K km s$^{-1}$, respectively.
  In the N$_2$H$^+$ map, black, purple, blue, green, and yellow contours show the integrated intensities at levels of 0.8, 1.6, 3.2, 4.8, and 6.4 K km s$^{-1}$, respectively.
  Red and yellow plus sings with labels show the dust clumps in L1688 (red: Motte et al.\ 1998; yellow: Johnstone et al.\ 2004) and L1689 (red: Nutter et al.\ 2006).
  Orange crosses with labels show four selected positions with C$_{2}$H emission peaks in the molecular ring Oph-RingSW.
  In the N$_{2}$H$^{+}$ integrated intensity map, to see these symbols clearly with a grey background, corsses are in black.
  Light blue plus signs with labels show our four newly named N$_2$H$^+$ clumps.
  Three dark green stars show the position of the protostar I16293 (left, 16$^{\rm h}$32$^{\rm m}$22.753$^{\rm s}$ -24\deg28$'$34.747$''$, J2000, in Kristensen et al.\ 2013),
  the star 22 Scorpius (bottom, 16$^{\rm h}$31$^{\rm m}$06.24$^{\rm s}$ -25\deg08$'$40.7$''$, J2000),
  and the star HD147889 (right, 16$^{\rm h}$26$^{\rm m}$17.74$^{\rm s}$ -24\deg29$'$47.9$''$, J2000).
  In L1688, red, yellow, and light blue letters A through M stand for sources Oph-A through M and
  orange numbers 1 through 4 stand for sources L1688-C$_{2}$H-1 through 4.
  In L1689, three red labels from north to south stand for L1689NW, L1689W and L1689S.
  In Section 3.1, the four N$_2$H$^+$ clumps with light blue crosses and labels are discussed.
  In Ssection 5.1, the C$_{2}$H and N$_2$H$^+$ spectra in the 14 positions with red plus sings (except G), orange crosses, or yellow plus signs (H only) are discussed.}
  \label{intg}
\end{figure}

\subsection[]{IRAS Temperature Map from COMPLETE}
  We downloaded the IRAS dust temperature map of the $\rho$-Oph MCC from the COMPLETE project webpage\footnote{https://www.cfa.harvard.edu/COMPLETE/index.html}.
  The IRAS dust temperature map was derived from infrared images in the latest release of IRAS data set (IRIS\footnote{http://irsa.ipac.caltech.edu/data/IRIS/}).
  The description of the method used to derive the IRAS dust temperature can be found in Schnee et al.\ (2005). 
  The IRAS dust temperature map of the whole $\rho$-Oph MCC region was published by Ridge et al. (2006).

  The sky coverage and pixel size of the IRAS dust temperature map differ from those of our C$_{2}$H and N$_2$H$^+$ spectral data cubes.
  An IDL code, $hastrom$ in the NASA IDL Astronomy User's Library\footnote{http://idlastro.gsfc.nasa.gov/},
  was used to crop and resample the temperature maps so that the sky coverage
  and pixel size of the processed map match those of our C$_{2}$H and N$_2$H$^+$ integrated intensity maps.
  The IRAS dust temperatures were used as proxies of the excitation temperatures.

\section[]{RESULTS}

\subsection{Integrated Intensity Maps of C$_{2}$H and N$_2$H$^+$}
  In L1688, eight 1.3 mm dust clumps (Oph-A through G, Motte et al.\ 1998, red plus signs with labels in Figure \ref{intg})
  and two 850 $\mu$m dust clumps (Oph-H and Oph-I, Johnstone et al.\ 2004,
  yellow plus signs with labels in Figure \ref{intg}) were all detected in N$_2$H$^+$ emission within one $\rm \sim$60$''$ beam.
  There is an extremely faint N$_2$H$^+$ clump towards the position of Oph-G with an averaged N$_{2}$H$^{+}$ integrated intensity of 0.8 K km s$^{-1}$ in a 3$\times$3 pixel region.
  After being smoothed by a 5$\times$5 pixel box, the integrated intensity here becomes lower than 0.8 K km s$^{-1}$ and thus no contours are seen there in Figure \ref{intg}.
  Four more faint N$_2$H$^+$ clumps in L1688 have been identified, namely
  Oph-J (the N$_2$H$^+$ peak at 16$^{\rm h}$28$^{\rm m}$24$^{\rm s}$\textbf{,} -24\deg36$'$12$''$, J2000),
  Oph-K (the N$_2$H$^+$ peak at 16$^{\rm h}$29$^{\rm m}$17$^{\rm s}$\textbf{,} -24\deg36$'$05$''$, J2000),
  Oph-L (the N$_2$H$^+$ peak at 16$^{\rm h}$29$^{\rm m}$03$^{\rm s}$\textbf{,} -24\deg40$'$45$''$, J2000),
  and Oph-M (the N$_2$H$^+$ peak at 16$^{\rm h}$28$^{\rm m}$47$^{\rm s}$ -24\deg45$'$04$''$, J2000).
  Oph-J is the brightest among these four clumps, with a 1.6 K km s$^{-1}$ peak N$_2$H$^+$ integrated intensity (4.2 $\sigma$).
  It was previously detected via 1.1 mm dust continuum (Bolo 26 in Young et al.\ 2006)
  and separated into four cores in the 1.2 mm dust continuum, namely MMS-60, 126, 130, and 99 in Stanke et al. (2006).
  Oph-L is the next brightest clump with a 1.2 K km s$^{-1}$ peak N$_2$H$^+$ integrated intensity (3.2 $\sigma$).
  At least three 1.2 mm dust cores (MMS-92, 109 and 122, Stanke et al.\ 2006) are embedded in Oph-L.
  Oph-K and Oph-M, each of which has 0.9 K km s$^{-1}$ peak N$_2$H$^+$ integrated intensity (2.3 $\sigma$).
  Oph-K was previously detected via 1.2 mm dust continuum and two cores were identified as MMS-102 and MMS-119 by Stanke et al.\ (2006). 
  Without previously known dust peaks, Oph-M is identified for the first time in this data set.

  We also identified a C$_2$H molecular ring, Oph-RingSW, in the relatively diffuse region southwest of L1688.
  Some parts of this ring can also be found in the COMPLETE $^{12}$CO map of L1688.
  The origin of this ring is unknown.

   In Section 5.2, we select the C$_2$H and N$_2$H$^+$ spectra at various peaks, perform the hyperfine fittings, compare the centroid velocities, line widths,
   and check if the N$_2$H$^+$ emission peak and the dust emission peak of the clumps Oph-A through F and H are located within one beam (60$''$).

\subsection{Column Densities and C$_{2}$H to N$_2$H$^+$ Ratios }

  Some positions in L1688 and L1689 have weak C$_{2}$H or N$_2$H$^+$ emission.
  The integrated intensities at these positions are significant, but the spectra themselves are too weak for hyperfine fitting, e.g., the C$_2$H spectrum in L1689NW (see Table 3).
  We assumed optically thin emission and followed the recipe below (see e.g. Li 2002)

  The column density of an upper level population of an optically thin transition is
  \begin{equation}
    \textit{$N_{u}^{0} = \frac{8 \pi k \nu^{2}}{h c^{3} A_{ul}} \int T_{s} d\upsilon$,}
    \label{equ_1}
  \end{equation}
  where,
  $\int T_{s} d\upsilon$ is the integrated intensity consisting of all six hyperfine components of the C$_{2}$H spectra or all seven hyperfine components of the N$_2$H$^+$ spectra.
  \textbf{In addition, }$k$ is the Boltzmann constant, $h$ is the Planck constant, $\nu$ is the frequency, and $c$ is the speed of light. 
  The total column density is
  \begin{equation}
    \textit{$N_{\rm tot}=F_{u} F_{\tau} F_{b} N^0_{u}$},
    \label{equ_2}
  \end{equation}
  where $F_{u}$, $F_{\tau}$, and $F_{b}$ are the level correction factor, the opacity correction factor, and the correction factor for the background, respectively.
  In this calculation, we only calculated the C$_2$H column densities at positions with a C$_2$H integrated intensity of 1.5 K km s$^{-1}$ or higher.
  The threshold for N$_2$H$^+$ column density calculation was 2 K km s$^{-1}$.
  The details can be found in Appendix B.

  We applied the opacity correction to N$_2$H$^+$ spectra with integrated intensities over [-8, 15] km s$^{-1}$ greater than 3.0 K km s$^{-1}$ and optical depths between 0.5 and 5.0.
  If the fitted optical depth was below 0.5 or above 5.0,
  however, noise may have made the result unreliable by skewing the peak temperature ratios between hyperfine components, depending on the signal-to-noise ratios of the components.
  Ignoring an optical depth of 0.5 will lead to an underestimation of by 27\%.
  A similar opacity correction was also done for C$_{2}$H spectra but for the spectra with integrated intensity greater than 1.0 K km s$^{-1}$.

  In our C$_{2}$H and N$_2$H$^+$ column density calculations,
  using IRAS dust temperatures as excitation temperatures of C$_{2}$H and N$_2$H$^+$ can cause large uncertainty.
  For example, the original IRAS temperatures have an 84$''$ pixel size and a 4.3$'$ spatial resolution (Schnee et al.\ 2005).
  This spatial resolution is approximately four times lower than our molecular line data.
  The IRAS temperatures come from far-infrared fluxes and may overestimate temperatures in compact, dense cores.
  Using the \textit{Herschel} Space Observatory 160 $\mu$m and 250 $\mu$m dust continuum maps and James Clerk Maxwell Telescope (JCMT) 450 $\mu$m and 850 $\mu$m dust continuum maps,
  Pattle et al.\ (2015) derived the mean line-of-sight dust temperatures of sources in the $\rho$-Oph MCC by fitting the spectral energy distribution (SED) with a modified blackbody distribution.
  These temperatures are lower than the IRAS temperatures.
  Lower temperatures result in lower C$_{2}$H and N$_2$H$^+$ column densities.
  For example, in the location of the source SM1 (see Motte et al.\ 1998, 16$^{\rm h}$26$^{\rm m}$27.36$^{\rm s}$, -24\deg23$'$52.8$''$, J2000),
  the C$_{2}$H and N$_2$H$^+$ integrated intensities are 12.3 K km s$^{-1}$ and 12.0 K km s$^{-1}$, respectively.
  Using the IRAS temperature (37 K), the column densities of C$_{2}$H and N$_2$H$^+$ are 6.8$\times$10$^{13}$ cm$^{-2}$ and 3.1$\times$10$^{12}$ cm$^{-2}$, respectively.
  These two column densities are both 4.6 times higher than those estimated with the temperature (17.2$\pm$0.6 K) in the same location in Pattle et al.\ (2015).

  The ratio between the column densities, however, is impervious to the temperature uncertainty for cloud conditions relevant here.
  We therefore use [C$_2$H]/[N$_2$H$^+$] to explore chemical evolution and physical properties of $\rho-$Oph MCC.
  [C$_2$H]/[N$_2$H$^+$] represents both the column density ratio of C$_2$H and N$_2$H$^+$ and their abundance ratio of C$_2$H and N$_2$H$^+$.
  Figure \ref{t_vary} shows an example of [C$_2$H]/[N$_2$H$^+$] changing with different temperature when the integrated intensities of C$_2$H or N$_2$H$^+$ are all assumed as 5 K km s$^{-1}$.
  The [C$_2$H]/[N$_2$H$^+$] values under different temperatures in Figure \ref{t_vary} come from the column density and abundance ratio calculations which are mentioned above and in Appendix A.
  For T$_{ex}$ $>$ 7 K, the [C$_2$H]/[N$_2$H$^+$] increases only slightly with increasing temperatures.
  Other than Figure \ref{t_vary},
  with different assumed C$_2$H or N$_2$H$^+$ integrated intensity,
  we also calculated the [C$_2$H]/[N$_2$H$^+$] values in different temperature.
  The weak dependence of abundance ratio on temperatures remains valid.
  In our observation region, all the sources in Pattle et al.\ (2015) have temperatures of 8.2 K (8.2$\pm$0.3 K for 88N SMM 1) or higher.
  The [C$_2$H]/[N$_2$H$^+$] is thus a robust probe of cloud conditions.

  Figure \ref{cmp1} shows the distribution of the measured abundance ratio in L1688 and L1689, which varies between 5 and 110.
  The differences in the distribution of the abundance ratio will be mentioned in Section 5.

\begin{figure}
\centering
  \includegraphics[width=120mm,angle=90]{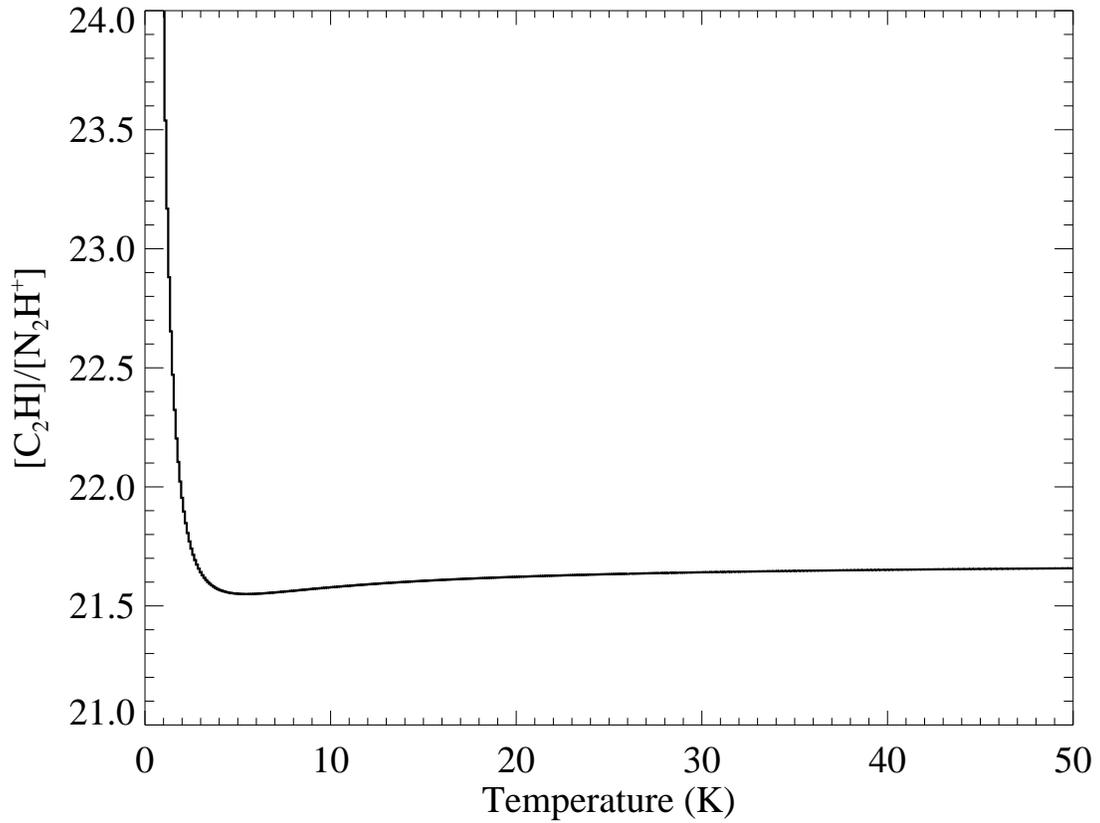}
  \caption{An example to show that [C$_2$H]/[N$_2$H$^+$] values are less affected by different temperatures.
  For the given C$_2$H and N$_2$H$^+$ integrated intensities (both are 5 K km s$^{-1}$ here), the values of [C$_2$H]/[N$_2$H$^+$] change little when the temperature is between $\rm \sim$7 K and 50 K. }
  \label{t_vary}
\end{figure}

\begin{figure}
\centering
\hspace{-5em}
\vspace{2em}
  \includegraphics[width=100mm,angle=90]{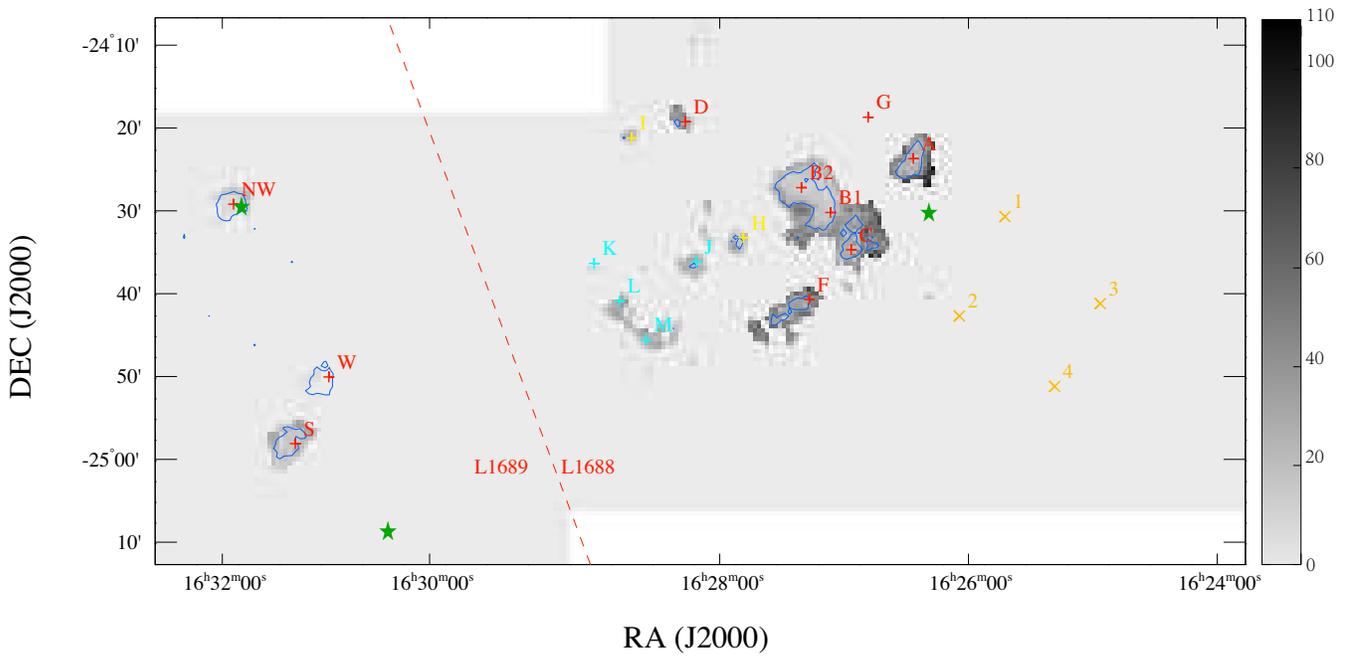}
  \caption{The distribution of [C$_2$H]/[N$_2$H$^+$].
  Blue contours represent the regions with 5 K km s$^{-1}$ or higher N$_2$H$^+$ integrated intensities.
  Symbols and labels are the same as those in Figure \ref{intg}.}
  \label{cmp1}
\end{figure}

\section[]{CHEMICAL MODEL}

  To interpret our data,
  we modeled [C$_{2}$H]/[N$_{2}$H$^{+}$] evolution with both the pure gas and gas-grain chemical networks.

  The simulation was performed using $Nautilus$, a gas-grain chemical code (Hersant et al.\ 2009),
  along with the gas-grain chemical reaction network from Hincelin et al.\ (2011) and Hincelin et al.\ (2013).
  An electronic version of the network is available in the KInetic Database for Astrochemistry (KIDA) database\footnote{http://kida.obs.u-bordeaux1.fr/models} ~\citep{Wakelam2012}.
  Low metal abundances, as in Semenov et al. (2010), had been assumed for the initial gas phase abundances.
  Initially, all species were in the gas phase.
  The cosmic-ray ionization rate was set to $1.3 \times 10^{-17}$ s$^{-1}$.
  The size of dust grains was assumed to be the same, with a radius of 0.1 $\mu$m.
  Both pure gas phase and gas-grain network simulation were carried out for comparative studies.
  Four different temperatures (10, 20, 30, and 40 K) and four different densities
  (5$\times$10$^{3}$ cm$^{-3}$,  2$\times$10$^{4}$ cm$^{-3}$, 1$\times$10$^{5}$ cm$^{-3}$, and 1$\times$10$^{6}$ cm$^{-3}$) were applied in our simulations.

  Our simulations are purely chemistry models without dynamic evolution.
  This is sufficient for examining the cloud chemistry as long as the time scale for density variation is much longer than the chemical relaxation time,
  which is a reasonable assumption for the densities concerned here.

  We compared the pure gas model and the gas-grain model.
  Figure \ref{model_1} shows the C$_{2}$H abundance,
  N$_2$H$^+$ abundance, and [C$_{2}$H]/[N$_{2}$H$^{+}$] as functions of time in a pure gas phase model
  and a gas-grain model with $n_{\rm H}$ = 1.0$\times$10$^{5}$ cm$^{-3}$ and $n_{\rm H}$ = 2.0$\times$10$^{4}$ cm$^{-3}$, respectively.
  The temperature in both models is fixed at 10 K.
  The most pronounced difference between these two models shows up in the abundance of C$_{2}$H.
  It remains constant or rises slightly in gas-grain models,
  while, in pure gas models, it plummets after approximately 10$^5$-10$^6$ years.
  The major destruction route of C$_{2}$H is reaction with O to form CO,
  while its major production route is the dissociative recombination of C$_2$H$_{2}^{+}$ and C$_2$H$_{3}^{+}$.
  Since only neutral species are allowed to accrete on grain surfaces in our gas-grain reaction network simulation,
  the abundances of C$_2$H$_{2}^{+}$ and C$_2$H$_{3}^{+}$ in the gas-grain model are not all that different from those in the pure gas model.
  O can be frozen on grain surfaces as H$_{2}$O and be largely depleted from the gas phase in later stages of the gas-grain model.
  C$_{2}$H experiences a lower destruction rate in gas-grain networks.
  Thus, the C$_{2}$H abundance in late stages is sensitive to the existence of dust mainly due to depletion of O onto grain surfaces.
  For N$_2$H$^+$, its major production route is the reaction
   \begin{equation}
   \textit{$\rm H^+_3 + N_2 \rightarrow N_2H^+ + H_2$}
   \label{reaction_1}
   \end{equation}
  while its major destruction route is reaction with CO (e.g., Turner 1995, Vasyunina et al.\ 2012).
  The depletion temperatures of both N$_2$ and CO are similar and lower than 20 K.
  Such temperatures result in that the trend of the change of N$_2$H$^+$ abundance is little affected by adding the dust into the model.
  So, the N$_2$H$^+$ abundance keeps rising in both the pure gas and gas-grain models, until t$\rm \sim$10$^{6}$ years.
  After t$\rm \sim$10$^{6}$ years, the decreasing of N$_2$H$^+$ abundance is the results of the evolution of CO, N$_2$, and H$^{+}_{3}$.
  Comparing the models with and without dust, the [C$_{2}$H]/[N$_{2}$H$^{+}$] values are insensitive to the existence of dust, also until t$\rm \sim$10$^{6}$ years.
  We use the gas-grain model in subsequent analyses.
  It is also worth noting that there is no significant difference seen in the evolution of C$_{2}$H abundances or N$_{2}$H$^{+}$ abundances between two densities that differ by a factor of 5.

\begin{figure}
\centering
  \includegraphics[width=160mm,angle=0]{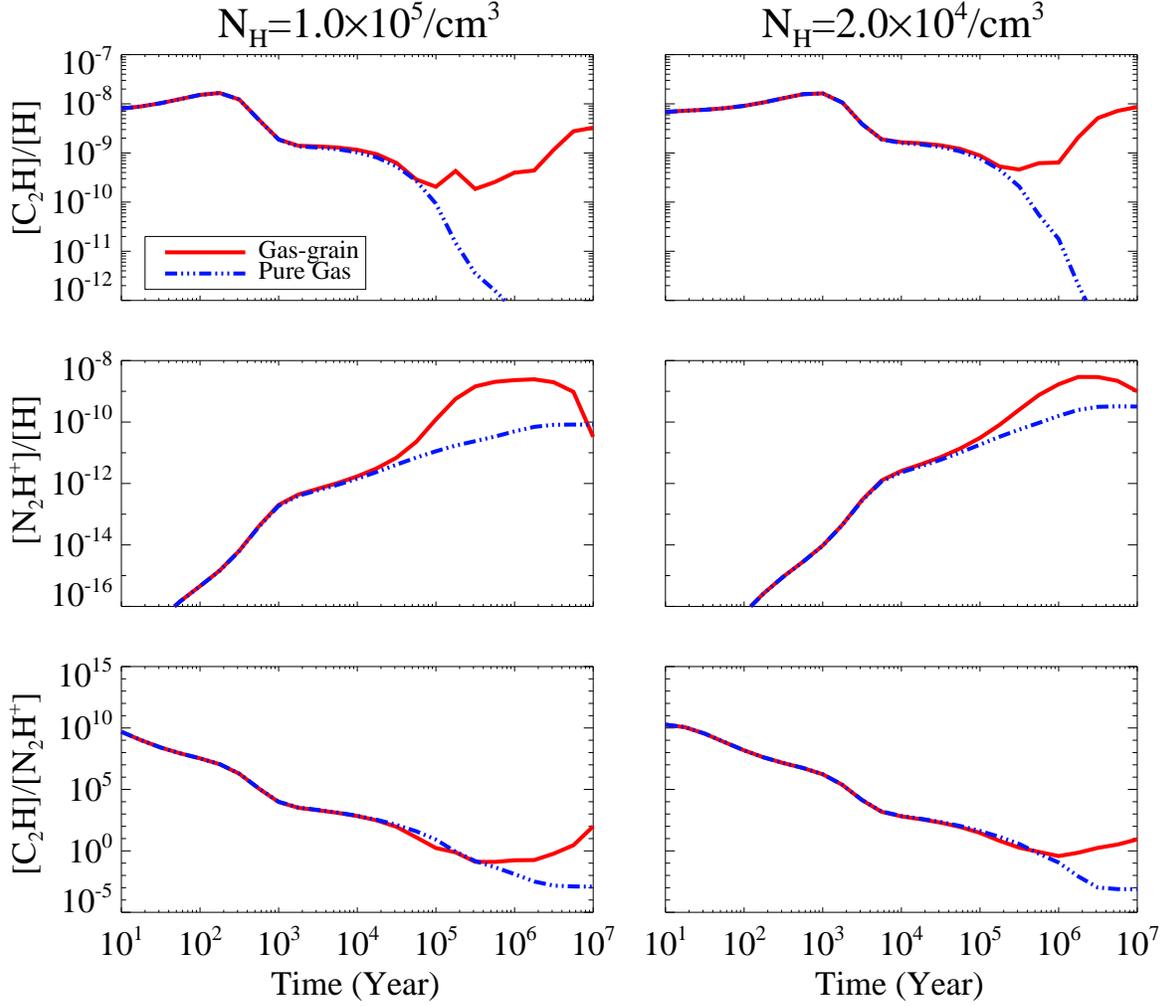}
  \caption{The results of pure gas and gas-grain models at a fixed temperature of 10 K.
  The red solid line shows the gas-grain model calculation of the C$_{2}$H abundance, the N$_2$H$^+$ abundance, and the [C$_{2}$H]/[N$_{2}$H$^{+}$].
  The blue dash-dotted line shows the results of pure gas model.
  }
  \label{model_1}
\end{figure}

  We ran the model with different extinctions and UV field strenghts to see if the variations of these two parameters affect the chemical evolution
  In the model, both the extinction and the UV field affect the photodissociation rate coefficients (hereafter PRCs), which are proportional to $F_0 G_0 \exp(-\gamma A_{\rm V})$.
  The factor $\gamma$ is a value around 2, depending on the balance between molecular photon absorption and differential dust absorption,
  $A$$_{\rm V}$ is the extinction,
  $G$$_{0}$ = 1.6$\times$10$^{-3}$ erg s$^{-1}$ cm$^{-2}$ is the unit of UV field, and $F_0$ is the scaling factor of the UV field.
  We ran the chemical model with two sets of parameter grids:
  1) $F_0$ = 1, density $n_{\rm H}$ = 2$\times$10$^{4}$ cm$^{-3}$, temperature T = 30 K, and extinctions $A_V$ = 1$^{\rm m}$, 5$^{\rm m}$, 8$^{\rm m}$, and 20$^{\rm m}$,
  and 2) $F_0$ = 0.01, 100, 1000, and 10000, density $n_{\rm H}$ = 2$\times$10$^{4}$ cm$^{-3}$, temperature T = 30 K, and extinction $A_V$ = 10$^{m}$.

  Different parts of molecular clouds may be exposed to different external FUV radiation fields,
  and different positions in molecular clouds have different extinctions.
  These variations produce different chemical products in different parts of a molecular cloud.
  The PRCs for cosmic ray-induced secondary photons, however, are independent of $A_{\rm V}$.
  For visual extinction A$_V$=10$^{\rm m}$,
  the PRCs due to secondary photons are more than three orders of magnitude larger than
  those of the external FUV radiation in our reaction network (Hincelin et al.\ 2011 and Hincelin et al.\ 2013).
  So, the variation of A$_V$ and the scaling factor do not affect the chemical evolution much in our models if A$_V$ is larger than 10.
  In our map, the extinction values in the place with C$_2$H or N$_2$H$^+$ emissions are relatively high (e.g.\textbf{,} 10 magnitude).
  We expect that the variation of extinction and external FUV radiation fields has little effect on the chemical evolution of dense molecular clouds..

  The extinctions at almost all the positions of the N$_2$H$^+$ clumps fulfill A$_{V}$ $\geq$ 10$^{\rm m}$.
  Figure \ref{vis} shows the evolutionary trends of the abundances of C$_2$H or N$_2$H$^+$ and their abundance ratio under different extinction.
  We see that the C$_2$H abundance, the N$_2$H$^+$ abundance, and the [C$_{2}$H]/[N$_{2}$H$^{+}$] all hardly differ for $A_V$ $\geq$ 8$^{\rm m}$,
  and the C$_{2}$H abundance is more easily affected by the different extinctions.
  The evolution of [C$_{2}$H]/[N$_{2}$H$^{+}$] with time differ substantially when $A_V$ = 1.
  Therefore, we ignored differences in extinctions and used 10$^{\rm m}$ as the basis for further model simulations.

  Figure \ref{fuv} shows the evolutionary trends of the abundances of C$_2$H or N$_2$H$^+$ and the abundance ratio under different external UV field.
  With extinction 10$^{\rm m}$,
  four models with $F_0$ values of 0.01, 100, 1000, and 10000 are shown.
  From the early evolutionary time of 10$^3$ years to $\rm \sim$5$\times$10$^5$ years for C$_{2}$H or 10$^3$ years to $\rm \sim$2$\times$10$^6$ years for N$_2$H$^+$,
  the model calculation results were little affected by the different $F_0$ values from 0.01 to 10000.
  From the later evolutionary time of $\rm \sim$5$\times$10$^5$ years to 10$^7$ years for C$_{2}$H or $\rm \sim$2$\times$10$^6$ years to 10$^7$ years for N$_2$H$^+$,
  the C$_{2}$H and N$_2$H$^+$ abundances only slightly increase up by factors of $\rm \sim$10 and 5, respectively.
  As seen from Figure \ref{vis}, when extinctions are high enough, both C$_{2}$H and N$_2$H$^+$ abundances are barely affected by the external UV field.
  For a cloud with extinction $A_V$ = 10$^{\rm m}$,
  however,
  the chemical evolution will be affected by the external UV field when $F_0$ reaches 1000.
  Liseau et al.\ (1999) estimated the UV field in the $\rho$-Oph MCC by sampling 33 positions, including Oph-A, Oph-B1, Oph-B2, Oph-C1 and Oph-D.
  The $F_0$ values (F$_{UV}^{\dagger}$ in Liseau et al.\ 1999) in these 33 positions are all below 150.
  Therefore, we ignored the differences in external UV field and used $F_0$ = 1 for further simulations.

\begin{figure}
\centering
  \includegraphics[width=160mm,angle=0]{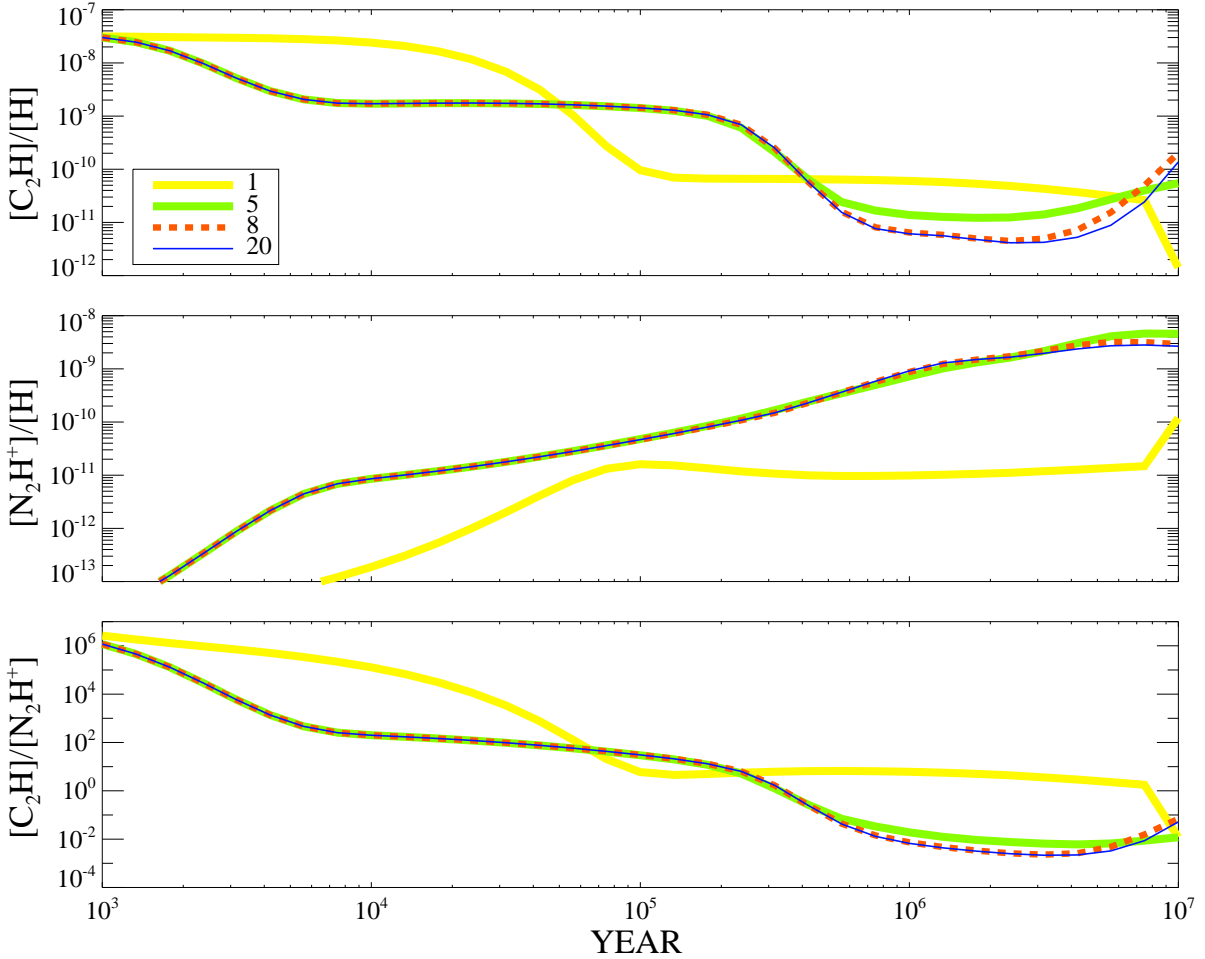}
  \caption{The gas-grain model calculation of C$_{2}$H abundance evolution of N$_2$H$^+$ and [C$_{2}$H]/[N$_{2}$H$^{+}$] under different extinctions.
  The yellow solid line, green solid line, red dashed line, and blue solid line show the results for extinctions $A_V$ = 1$^{\rm m}$, 5$^{\rm m}$, 8$^{\rm m}$, and 20$^{\rm m}$, respectively.
  F$_0$ = 1 was used for all cases.
  }
  \label{vis}
\end{figure}

\begin{figure}
\centering
  \includegraphics[width=160mm,angle=0]{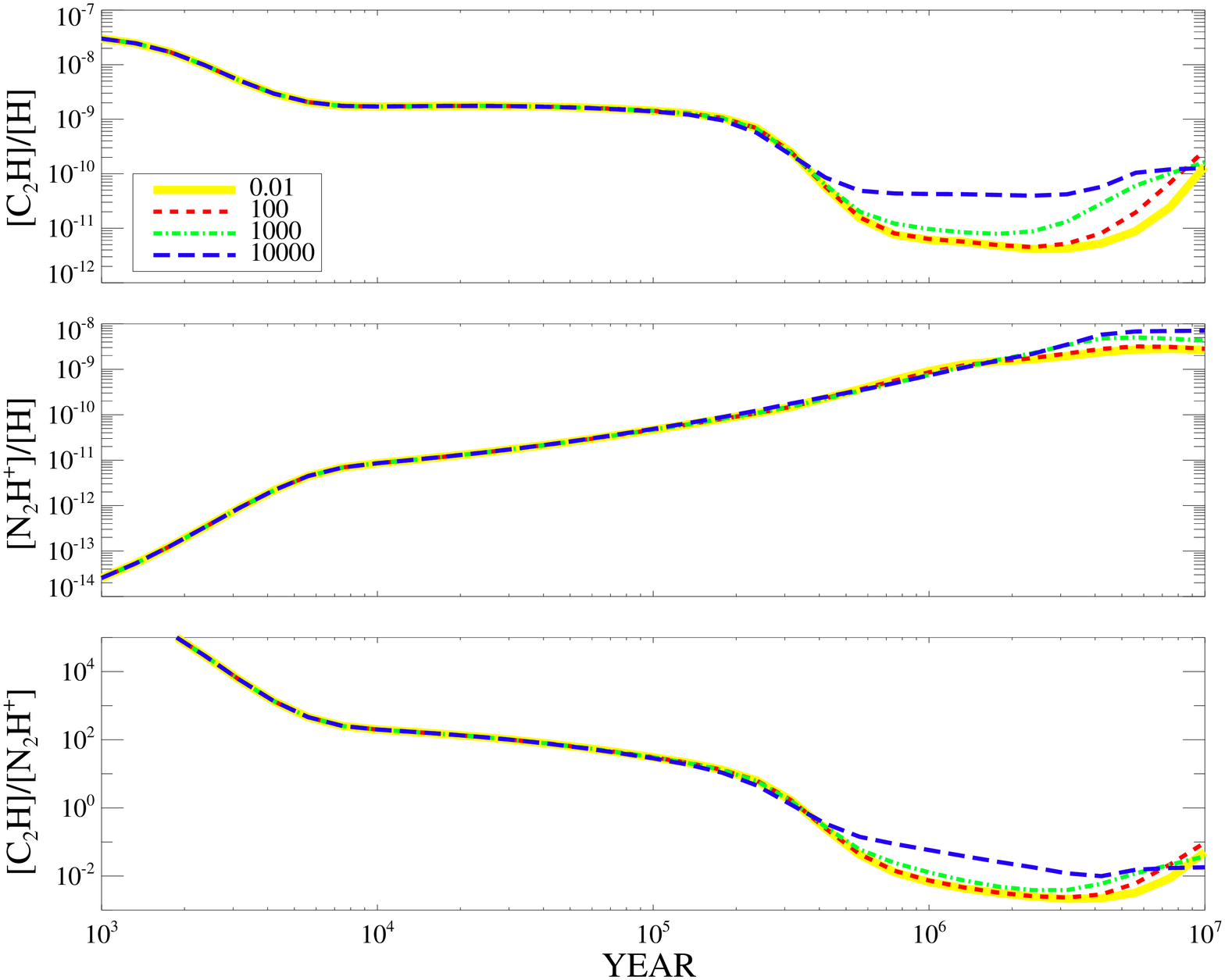}
  \caption{The gas-grain model calculation of C$_{2}$H abundance, N$_2$H$^+$ abundance, and [C$_{2}$H]/[N$_{2}$H$^{+}$] with different values of $F_0$.
  The yellow solid line, red dashed line, green dash-dotted line, and blue dashed line show the results of $F_0$ of 0.01, 100, 1000, and 10000, respectively.
  $A_V$=$10^m$ was used for all cases.
  }
  \label{fuv}
\end{figure}

  Figure \ref{model} shows the abundances of C$_{2}$H and
  N$_2$H$^+$ and [C$_{2}$H]/[N$_{2}$H$^{+}$] as functions of time for different temperatures and densities in the gas-grain model.
  Here, we find that the C$_{2}$H abundance depends strongly on temperature and density.
  These dependencies can be explained from the formation and destruction routes of C$_{2}$H.
  The abundance of C$^{+}$ plays a crucial role in the production of C$_{2}$H.
  Protonated methane is produced by radiative association of C$^+$ with H$_2$ and hydration addition reactions.
  All carbon atoms are in the form of C$^{+}$ in the beginning of the model and are gradually converted to species such as CO or protonated methane.
  Thus, the abundance of C$_{2}$H increases in the beginning when abundances of methane or species such as CO and OH also increase.
  Atomic O accretes on grain surfaces and reacts with other species, and thus is depleted from the gas phase.
  At 10 K, when most gas phase O atoms are depleted after approximately 2$\times$10$^{6}$ years, C$_{2}$H can be hardly destroyed and the abundance of C$_{2}$H increases again.
  Indeed, our simulation results bear out this trend.
  O starts to desorb from grain surfaces at around 15 K.
  As temperature increases, it becomes increasingly difficult to freeze atomic O on grain surfaces.
  Hence, C$_{2}$H abundance decreases with increasing temperature at late stages.
  Density also affects C$_{2}$H abundance.
  The dependence of C$_{2}$H abundance on density is due to both O and C$^+$ are depleting more quickly from the gas phase when density increases.
  The desorption of O at around 15 K is also the reason for that the 20 K models differ so dramatically from those at lower and higher temperatures.
  The situation is different for the changes of N$_2$H$^+$ abundances.
  The desorption energies are 1000 K for N$_2$ and 1150 K for CO. 
  Both N$_2$ and CO can be depleted from the gas phase when the temperature is lower than 20 K.
  The relative abundance of these two species determines the abundance of N$_2$H$^+$.
  The N$_2$H$^+$ abundance is affected little by temperature and volume density variations.

  Combining the evolutions of both species, the
  [C$_{2}$H]/[N$_{2}$H$^{+}$] decreases with time between 10$^2$ years and 10$^6$ years
  for a volume density of 5.0$\times$10$^3$ cm$^{-3}$ and from the beginning to approximately 2.0$\times$10$^4$ years for a volume density of 1.0$\times$10$^6$ cm$^{-3}$.
  This evolutionary trend is little affected by temperature for T$_k$ $<$ 40 K.
  At high density (n$_H$ $>$ 10$^5$ cm$^{-3}$),
  the time it takes for the abundance ratio to drop at least one order of magnitude becomes less than the typical dynamical time.
  For example, a 0.1 pc clump at the distance of $\rho$-Oph MCC spans about 2.9$'$, implying a turbulence crossing time of $\rm \sim$10$^5$ year for a characteristic 1 km s$^{-1}$ turbulence velocity.

  Concerning the observed [C$_{2}$H]/[N$_{2}$H$^{+}$] values, the trends for all three modeled densities are consistently decreasing with time.
  The possible volume density in the positions with the observed [C$_{2}$H]/[N$_{2}$H$^{+}$] values are 1.0$\times$10$^5$ cm$^{-3}$ or higher (e.g., see Motte et al.\ 1998, Young et al.\ 2006).
  In such codidtions, the [C$_{2}$H]/[N$_{2}$H$^{+}$] values are more sensitive to the volume densities than to the chemical ages, as we describe below.

\begin{figure}
\centering
  \includegraphics[width=160mm,angle=0]{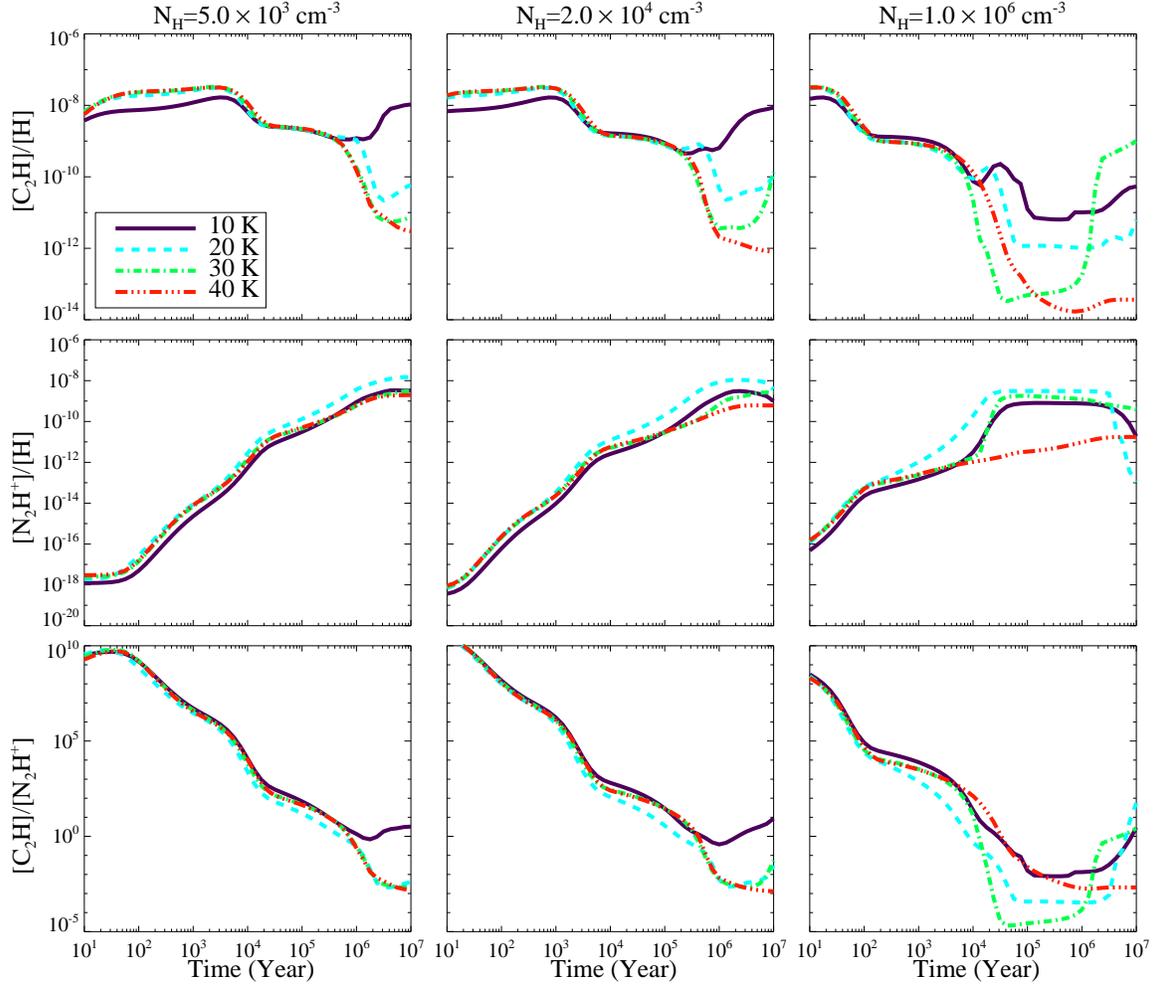}
  \caption{The gas-grain model calculation of C$_{2}$H abundance, N$_2$H$^+$ abundance and [C$_{2}$H]/[N$_{2}$H$^{+}$].
  Each column shows the result for the same density. The densities are shown as the titles of the first plots of the column.
  The dark purple solid line, light blue dash line, light green dash-dot line, and red dash-dotted line show the calculation of the gas-grain model with 10 K, 20 K, 30 K and 40 K, respectively.
  }
  \label{model}
\end{figure}

\section{DISCUSSION}

\subsection{[C$_{\rm 2}$H]/[N$_{\rm 2}$H$^{\rm +}$] as a Tracer of Chemical Evolution and Volume Density}

  According to our chemical model, the C$_{2}$H abundance decreases while the N$_{2}$H$^{+}$ abundance increases with time (see Figure \ref{model}).
  Therefore, [C$_{\rm 2}$H]/[N$_{\rm 2}$H$^{\rm +}$] can be used as a tracer of chemical evolution.
  Figure \ref{N_chemi} shows a comparison between the observed [C$_{\rm 2}$H]/[N$_{\rm 2}$H$^{\rm +}$] values in L1688 or L1689
  and models of different volume density and temperature.
  The mean estimated chemical ages of the observed [C$_{\rm 2}$H]/[N$_{\rm 2}$H$^{\rm +}$] values in L1688 (yellow squares in Figure \ref{N_chemi}) or
  L1689 (purple triangles in Figure \ref{N_chemi}) are listed in Table \ref{chemi_age}.
  Linear interpolation is used to include our observed [C$_{\rm 2}$H]/[N$_{\rm 2}$H$^{\rm +}$] values with the modeled data.
  The details of the interpolation can be found in Appendix C.
  Though the IRAS temperatures could also bring uncertainties here, as shown in Figure \ref{N_chemi}, the uncertainties brought about by temperature are insignificant.

  Figure \ref{N_chemi} shows that both chemical ages and volume densities affect the [C$_{\rm 2}$H]/[N$_{\rm 2}$H$^{\rm +}$] values.
  In the first and second panels from the left of Figure \ref{N_chemi},
  the chemical ages of the observed [C$_{\rm 2}$H]/[N$_{\rm 2}$H$^{\rm +}$] are both around 10$^5$ years.
  In such a volume density range, the estimated chemical ages seem to be affected more by the [C$_{2}$H]/[N$_{2}$H$^{+}$] values rather than by different volume densities.
  From left to right, in the third and fourth panels in Figure \ref{N_chemi},
  for these same observed [C$_{\rm 2}$H]/[N$_{\rm 2}$H$^{\rm +}$] values,
  the estimated ages decrease to $\rm \sim$5$\times$10$^{4}$ years and $\rm \sim$1$\times$10$^{4}$ years, respectively.
  For the same [C$_{\rm 2}$H]/[N$_{\rm 2}$H$^{\rm +}$] values and in high volume density gas,
  Figure \ref{N_chemi} shows that higher volume densities result in younger estimated ages of the same observed abundance ratios,
  and both the [C$_{2}$H]/[N$_{2}$H$^{+}$] values and volume densities affect the chemical ages at similar levels.
  In addition, the relationship between the time/chemical age and [C$_{2}$H]/[N$_{2}$H$^{+}$] steepens when the volume density increases.
  The mean ages in Table \ref{chemi_age} corresponding to the observed [C$_{\rm 2}$H]/[N$_{\rm 2}$H$^{\rm +}$] values in L1688 or L1689
  decrease more when the volume density increases to higher volume density ranges.
  For example, when the volume density increases 20 times from 5.0$\times$10$^{3}$ cm$^{-3}$ to 1$\times$10$^{5}$ cm$^{-3}$,
  the derived chemical mean age corresponding to the observed abundance values in L1688 decreases from 1.4$\times$10$^5$ years to one-third of this value.
  When the volume density increases 10 times further from 10$^{5}$ cm$^{-3}$,
  the derived chemical mean age corresponding to the observed abundance values in L1688 decreases from 4.4$\times$10$^4$ year to one-fifth of that value.

  \begin{figure}
  \centering
      \includegraphics[width=180mm,angle=0]{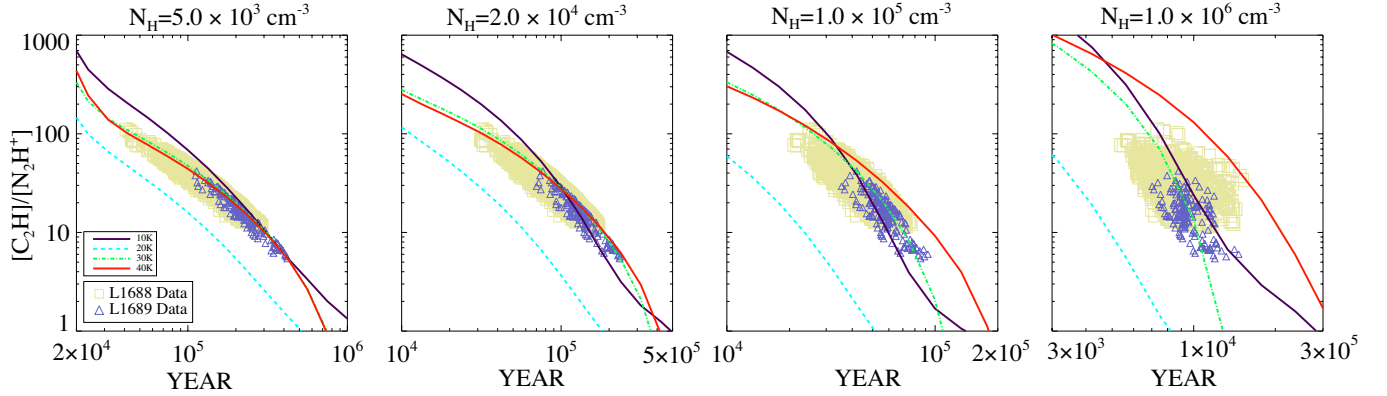}
      \caption{The modeled [C$_{2}$H]/[N$_{2}$H$^{+}$] (lines) as a function of evolutionary time with the real observation data (squares and triangles).
      In each plot, the dark purple solid line, light blue dashed line, light green dash-dotted line, and red dash-dotted line
      show the calculation results of the gas-grain model with temperatures of 10 K, 20 K, 30 K, and 40 K, respectively.
      Squares and triangles show observed [C$_{2}$H]/[N$_{2}$H$^{+}$] values of each pixel in L1688 and L1689, respectively.
      In this Figure, the modeled data we use are as same as these in Figure \ref{model}.
      To show the observed data points clearly, we use the different scales on both axes.
      }
    \label{N_chemi}
  \end{figure}

  \begin{table}
\centering
  \caption{Mean chemical ages}
\vspace{1em}
  \begin{tabular}{lcc}
  \hline
  Density $N_{H}$      &  L1688 Mean Age      &  L1689 Mean Age   \\
  ($$ cm$^{-3}$)       &    (yr)                &   (yr)             \\
  \hline
  $5.0 \times 10^{3}$  &   $1.4 \times 10^{5}$   &  $2.3 \times 10^{5}$   \\
  $2.0 \times 10^{4}$  &   $8.7 \times 10^{4}$   &  $1.4 \times 10^{5}$   \\
  $1.0 \times 10^{5}$  &   $4.4 \times 10^{4}$   &  $5.8 \times 10^{4}$   \\
  $1.0 \times 10^{6}$  &   $8.7 \times 10^{3}$   &  $9.7 \times 10^{3}$   \\
  \hline
  \label{chemi_age}
\end{tabular}
\end{table}

  Figure \ref{chemical_age_n} in another way shows the relationship between [C$_{2}$H]/[N$_{2}$H$^{+}$], the chemical age, and the volume density.
  The top panel shows using models how [C$_{2}$H]/[N$_{2}$H$^{+}$] is affected by chemical age.
  When the volume density is low, e.g., 5$\times$10$^{3}$ cm$^{-3}$,
  the abundance ratios at different ages look distinct in the panel.
  When volume density is high, e.g. 10$^6$ cm$^{-3}$, the abundance ratios at different ages seem to be indistinct in the panel.
  Note that the axis for chemical age (Y axis) is linear to reflect the chemical age decreasing when the volume density is increasing,
  thus, the total range variation at high density is hard to recognize in such a panel, although they may still significant.
  The bottom panel shows how [C$_{2}$H]/[N$_{2}$H$^{+}$] is affected by volume density.
  When the volume density is 10$^5$ cm$^{-3}$ or lower,
  chemical age changes of one order of magnitude higher or lower result in [C$_{2}$H]/[N$_{2}$H$^{+}$] changes of only $\rm \sim$1.5 orders of magnitude lower or higher, respectively.
  When the volume density is 10$^6$ cm$^{-3}$,
  however, [C$_{2}$H]/[N$_{2}$H$^{+}$] drops $\rm \sim$7 orders of magnitude lower when the chemical age only increases one order of magnitude higher.
  It appears that 10$^5$ cm$^{-3}$ is roughly the critical value between the regimes where [C$_{2}$H]/[N$_{2}$H$^{+}$] is sensitive to volume density or not.

  From our observations, [C$_{2}$H]/[N$_{2}$H$^{+}$] varies from 5 to 110.
  In the top panel of Figure \ref{chemical_age_n}, two non-vertical dotted lines show the results of the [C$_{2}$H]/[N$_{2}$H$^{+}$] values of 5 and 110.
  Motte et al.\ (1998) evaluated the volume densities of cores
  in L1688 with their 1.3 mm dust continuum map, finding values between 1.4$\times$10$^{5}$ cm$^{-3}$ and 8.0$\times$10$^{6}$ cm$^{-3}$.
  These two volume densities are also shown in the top panel of Figure \ref{chemical_age_n} with two vertical dotted lines.
  For comparison, Young et al.\ (2006) evaluated the volume densities of cores with their 1.1 mm dust continuum map which included both L1688 and L1689.
  Using a method different from the one used by Motte et al.\ (1998), they found 48 cores in their map.
  Among these 48 cores, 41 are in L1688 or L1689 and these have volume densities between 1$\times$10$^{5}$ cm$^{-3}$ and 3$\times$10$^{7}$ cm$^{-3}$ which do not differ too much from the results of Motte et al.\ (1998).
  With such limits of observed [C$_{2}$H]/[N$_{2}$H$^{+}$] values and volume densities (the region with dotted line boundries),
  for the overall values of [C$_{2}$H]/[N$_{2}$H$^{+}$], they are likely the results from a combination of chemical age and volume density.

  In the integrated intensity map (see Figure \ref{intg}),
  the C$_2$H emission is more extended in the relatively low density regions (e.g., the C$_2$H ring) in L1688 than in L1689.
  The emission distribution difference can be explained by L1688 having chemically younger gas in relatively less dense regions which are without detectable N$_2$H$^+$ emission.
  With no N$_2$H$^+$ detection, the values of [C$_{2}$H]/[N$_{2}$H$^{+}$] indicate that these regions have young chemical ages.
  In the abundance ratio map (see Figure \ref{cmp1}),
  the observed [C$_{2}$H]/[N$_{2}$H$^{+}$] values in L1688 tend to be higher than those in L1689.
  For any given clump, the abundance ratio in the inner region tends to be lower than in the outer region.
  For the overall distribution of [C$_{2}$H]/[N$_{2}$H$^{+}$],
  they are also likely the results of time evolution, accelerated at higher densities.

  \begin{figure}
  \centering
      \includegraphics[width=130mm,angle=0]{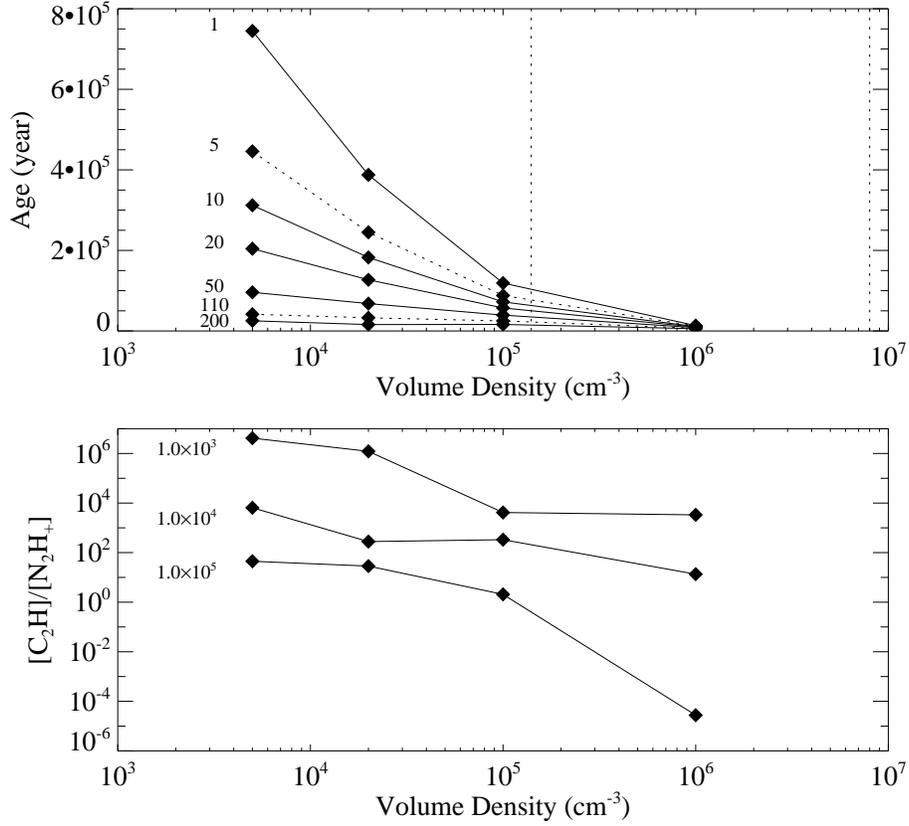}
      \caption{
      The modeled [C$_{2}$H]/[N$_{2}$H$^{+}$] values vary with chemical ages (upper) or different volume densities (lower) when $T_k$ = 30 K.
      In the upper panel, the solid lines and non-vertical dotted lines show how the estimated ages change with one [C$_{2}$H]/[N$_{2}$H$^{+}$] of which the value is shown at the left of the line.
      The two non-vertical dotted lines show the minimum and maximum values of the observed [C$_{2}$H]/[N$_{2}$H$^{+}$] values.
      The two vertical dotted lines show the positions with volume densities of 1.4$\times$10$^{5}$ cm$^{-3}$ to 8.0$\times$10$^{6}$ cm$^{-3}$, respectively.
      In the lower panel, each line shows how the [C$_{2}$H]/[N$_{2}$H$^{+}$] values change with one chemical age of which the value is shown at the left of the line.
      In these two panels, filled diamonds show the positions of data points. }
    \label{chemical_age_n}
  \end{figure}

\subsection{Centroid Velocities and Line-widths Comparison between C$_{2}$H and N$_2$H$^+$}

  We selected 14 positions of C$_{2}$H or N$_{2}$H$^{+}$ intensity peaks in our maps and analysed their spectra.
  Figure \ref{spec} shows spectra from these positions. The locations of these positions are represented by crosses and X letters in Figure \ref{intg}.
  The coordinates, RMS values, centroid velocities with errors, and line widths with errors of these spectra are listed in Table \ref{lines}.
  We ignored Oph-E, because it is only 5$'$ south of Oph-C and connected in N$_{2}$H$^{+}$ emission.
  The eight N$_2$H$^+$ peaks in clumps of L1688 (Oph-A through F and H) are all co-located
  with dust continuum sources (e.g., dust clumps in Motte et al. 1998), i.e., within 42$''$ or less.

\begin{table}
\rotate
\begin{tiny}
\caption{Lines hyperfine fitting result. $CLASS$ fitting errors of centroid velocities and line widths are shown in the brackets behind the numbers.}
\vspace{1em}
\hspace{-5em}
  \begin{tabular}{lcccccccc}
                   \hline
  Source           & Molecular   & RA            &  DEC       & RMS    & V               &$\Delta$ v     & Offset from N$_2$H$^+$ Peak            &  Comment   \\
                   &             & (J2000)       & (J2000)    & (K)    &(km s$^{-1}$)    & (km s$^{-1}$) & ($''$ and direction)$^*$ &          \\
                   \hline
  Oph-A            &  N$_2$H$^+$     &  16$^{\rm h}$26$^{\rm m}$28.4$^{\rm s}$  &  -24\deg23$'$42$''$  & 0.27   &  3.58(0.02)           &  0.79(0.05)         &  30$''$  E                    &  SM1$^{a}$ in Oph-A   \\
                   &  C$_{2}$H       &  16$^{\rm h}$26$^{\rm m}$28.2$^{\rm s}$  &  -24\deg23$'$42$''$  & 0.26   &  3.39(0.03)           &  1.11(0.05)         &                                  &    \\
                   \hline
  Oph-B1           &  N$_2$H$^+$     &  16$^{\rm h}$27$^{\rm m}$12.4$^{\rm s}$  &  -24\deg30$'$12$''$  & 0.26   &  3.75(0.02)           &  0.96(0.08)         &  30$''$  W                    &  MM3$^{a}$ in Oph-B1     \\
                   &  C$_{2}$H       &  16$^{\rm h}$27$^{\rm m}$12.2$^{\rm s}$  &  -24\deg30$'$12$''$  & 0.27   &  3.61(0.05)           &  1.18(0.20)         &                                  &      \\
                   \hline
  Oph-B2           &  N$_2$H$^+$     &  16$^{\rm h}$27$^{\rm m}$27.7$^{\rm s}$  &  -24\deg27$'$12$''$  & 0.24   &  4.03(0.01)           &  1.07(0.03)         &  0$''$                      &  MM8$^{a}$ in Oph-B2  \\
                   &  C$_{2}$H       &  16$^{\rm h}$27$^{\rm m}$27.5$^{\rm s}$  &  -24\deg27$'$12$''$  & 0.27   &  4.02(0.05)           &  1.06(0.12)         &                                  &      \\
                   \hline
  Oph-C            &  N$_2$H$^+$     &  16$^{\rm h}$27$^{\rm m}$01.4$^{\rm s}$  &  -24\deg34$'$42$''$  & 0.26   &  3.72(0.02)           &  0.56(0.04)         &   42$''$  SE                  &  MM6$^{a}$ in Oph-C       \\
                   &  C$_{2}$H       &  16$^{\rm h}$27$^{\rm m}$01.2$^{\rm s}$  &  -24\deg34$'$42$''$  & 0.28   &  3.78(0.03)           &  0.71(0.04)         &                                  &        \\
                   \hline
  Oph-D            &  N$_2$H$^+$     &  16$^{\rm h}$28$^{\rm m}$29.1$^{\rm s}$  &  -24\deg19$'$12$''$  & 0.28   &  3.47(0.02)$^{\#}$    &  0.44(0.04)$^{\#}$  &   42$''$  SE                  &  MM2$^{a}$ in Oph-D    \\
                   &  C$_{2}$H       &  16$^{\rm h}$28$^{\rm m}$28.9$^{\rm s}$  &  -24\deg19$'$12$''$  & 0.31   &  3.49(0.02)           &  0.71(0.06)         &                                  &           \\
                   \hline
  Oph-F            &  N$_2$H$^+$     &  16$^{\rm h}$27$^{\rm m}$23.4$^{\rm s}$  &  -24\deg40$'$42$''$  & 0.26   &  4.13(0.03)           &  0.82(0.08)         &   30$''$  W                   &  MM2$^{a}$ in Oph-F    \\
                   &  C$_{2}$H       &  16$^{\rm h}$27$^{\rm m}$23.3$^{\rm s}$  &  -24\deg40$'$42$''$  & 0.29   &  4.23(0.04)           &  1.09(0.12)         &                                  &        \\
                   \hline
  Oph-H            &  N$_2$H$^+$     &  16$^{\rm h}$27$^{\rm m}$58.5$^{\rm s}$  &  -24\deg33$'$12$''$  & 0.30   &  4.18(0.03)           &  0.63(0.07)         &    30$''$  S                  &  MM1$^{c}$ in Oph-H \\
                   &  C$_{2}$H       &  16$^{\rm h}$27$^{\rm m}$58.4$^{\rm s}$  &  -24\deg33$'$12$''$  & 0.29   &  4.24(0.02)           &  0.58(0.02)         &                                  &          \\
                   \hline
  L1688-C$_{2}$H-1 &  C$_{2}$H       &  16$^{\rm h}$25$^{\rm m}$39.9$^{\rm s}$  &  -24\deg30$'$42$''$  & 0.24   &  3.56(0.08)           &  2.08(0.18)         &                                   &          \\
  L1688-C$_{2}$H-2 &  C$_{2}$H       &  16$^{\rm h}$26$^{\rm m}$04.0$^{\rm s}$  &  -24\deg42$'$42$''$  & 0.27   &  3.45(0.10)           &  1.37(0.10)         &                                   &          \\
  L1688-C$_{2}$H-3 &  C$_{2}$H       &  16$^{\rm h}$24$^{\rm m}$49.2$^{\rm s}$  &  -24\deg41$'$42$''$  & 0.35   &  3.00(0.15)           &  2.46(0.38)         &                                   &          \\
  L1688-C$_{2}$H-4 &  C$_{2}$H       &  16$^{\rm h}$25$^{\rm m}$13.3$^{\rm s}$  &  -24\deg51$'$42$''$  & 0.31   &  3.28(0.10)           &  1.59(0.29)         &                                   &          \\
                   \hline
  L1689W           &  N$_2$H$^+$     &  16$^{\rm h}$31$^{\rm m}$39.1$^{\rm s}$  &  -24\deg49$'$42$''$  & 0.31   &  4.46(0.02)           &  0.59(0.04)         &   42$''$  SW                  &  SMM16$^{b}$ in L1689$-$West      \\
                   &  C$_{2}$H       &  16$^{\rm h}$31$^{\rm m}$39.0$^{\rm s}$  &  -24\deg49$'$42$''$  & 0.33   &  4.57(0.09)           &  0.79(0.19)         &                                  &        \\
                   \hline
  L1689S           &  N$_2$H$^+$     &  16$^{\rm h}$31$^{\rm m}$57.1$^{\rm s}$  &  -24\deg57$'$42$''$  & 0.27   &  4.41(0.02)           &  0.62(0.03)         &   42$''$  SW                  &  SMM8$^{b}$ in L1689$-$South         \\
                   &  C$_{2}$H       &  16$^{\rm h}$31$^{\rm m}$57.0$^{\rm s}$  &  -24\deg57$'$42$''$  & 0.32   &  4.47(0.05)           &  1.20(0.16)         &                                  &            \\
                   \hline
  L1689NW          &  N$_2$H$^+$     &  16$^{\rm h}$32$^{\rm m}$28.8$^{\rm s}$  &  -24\deg28$'$42$''$  & 0.40   &  3.76(0.02)           &  0.62(0.04)         &    32$''$  S                 &  SMM19$^{b}$ in L1689$-$NorthWest   \\
                   &  C$_{2}$H       &  16$^{\rm h}$32$^{\rm m}$28.6$^{\rm s}$  &  -24\deg28$'$42$''$  & 0.37   &  -                    &  -            &                                  &  weak C$_{2}$H emission     \\
                   \hline
  \tablenotetext{a}{  1.3 mm dust continuum sources in Motte et al.\ (1998). }
  \tablenotetext{b}{  850 $\mu$m dust continuum sources in Nutter et al.\ (2006). }
  \tablenotetext{c}{  850 $\mu$m dust continuum sources in Johnstone et al.\ (2004). }
  \tablenotetext{*}{  The numbers in this column are the angular distances between the positions of the dust sources and the peaks of the N$_2$H$^+$ emission in the integrated intensity map. }
  \tablenotetext{\#}{  The Gildas/CLASS reports that it is an optimistic fitting. }
  \tablenotetext{-}{  The line is too weak to fit the hyperfine structures. }
          \label{lines}
        \end{tabular}
    \end{tiny}
\end{table}

\begin{figure}
\centering
  \includegraphics[width=150mm,angle=0]{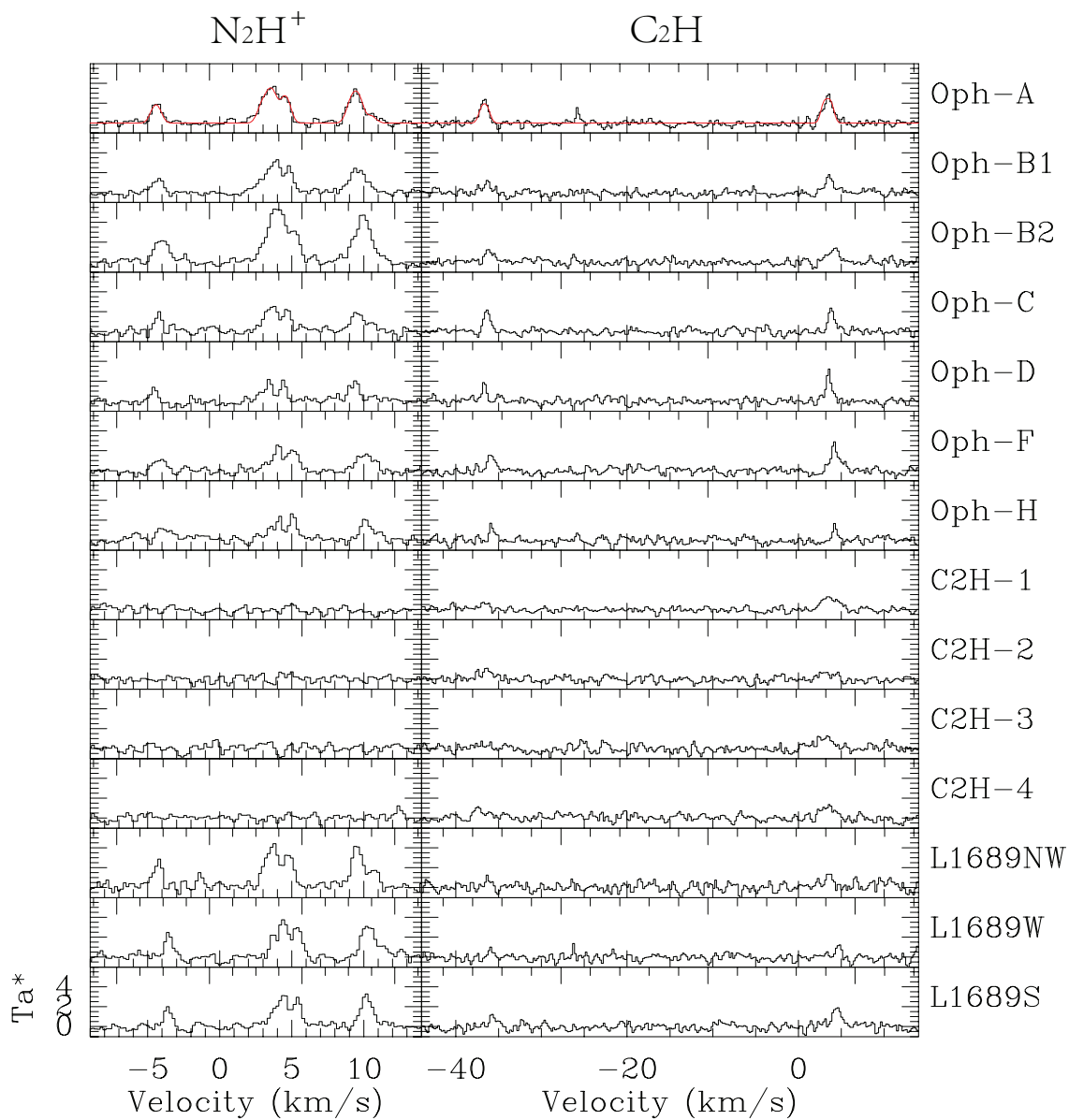}
  \caption{The N$_2$H$^+$ and C$_{2}$H spectra in the selected 14 positions.
  Each row is the N$_2$H$^+$(left) and C$_{2}$H(right) spectra with the source name in the right of the C$_{2}$H spectrum.
  The red lines show the hyperfine structure fitting results of Oph-A's N$_2$H$^+$ and C$_{2}$H spectra.
  In some C$_{2}$H spectra, the peaks near -25 km s$^{-1}$ are bad channels. }
  \label{spec}
\end{figure}

  L1688-C$_2$H-1, L1688-C$_2$H-2, L1688-C$_2$H-3, and L1688-C$_2$H-4 are located in the molecular ring Oph-RingSW.
  Their line widths are larger than those at other C$_{2}$H peak positions in L1688 or L1689.
  The centroid velocities of L1688-C$_2$H-3 and L1688-C$_2$H-4 are smaller than those of any other C$_{2}$H spectra at the selected positions.
  In these four positions, no N$_2$H$^+$ emission is detected.
  With small centroid velocities and large line widths,
  the C$_{2}$H spectra at these four positions indicate that the molecular ring is not a part of the denser regions of L1688.
  In L1689NW, the C$_{2}$H line is detectable but weak.
  For the other nine selected sources with both C$_{2}$H and N$_{2}$H$^{+}$ spectra, the centroid velocity differences between C$_{2}$H and N$_{2}$H$^{+}$ spectra are smaller than 0.19 km s$^{-1}$,
  while the velocity resolutions of C$_{2}$H and N$_{2}$H$^{+}$ spectra are 0.21 km s$^{-1}$ and 0.20 km s$^{-1}$, respectively.
  In these nine positions, differences in line widths between C$_{2}$H and N$_{2}$H$^{+}$ spectra are smaller than 0.27 km s$^{-1}$, except for these of Oph-A (0.32 km s$^{-1}$) and L1689S (0.58 km s$^{-1}$).
  Such similar centroid velocities and line widths indicate that C$_{2}$H N=1-0 and N$_{2}$H$^{+}$ J=1-0 trace similar gas in our observation region.
  Although the differences in both the centroid velocities and the line widths are similar to the channel width, from the fitting errors,
  the line widths of C$_{2}$H are significantly larger than those of N$_{2}$H$^{+}$ in almost all cases, except in the positions of Oph-B2 and Oph-H.
  This behaviour confirms our understanding that C$_2$H traces more extended and relatively more diffuse gas than N$_2$H$^+$ despite their transitions having similar critical densities.

\section{CONCLUSIONS}

  We have mapped 2.5 square degrees of the $\rho$ Ophiuchi Molecular Cloud Complex in the emission of the C$_{2}$H N=1-0 and N$_2$H$^+$ J=1-0 transitions.
  The observations are the first large-scale C$_{2}$H and N$_2$H$^+$ maps of the $\rho$-Oph MCC.
  Our main results are:

  1) Four N$_2$H$^+$ clumps and one C$_{2}$H ring are identified and named here as Oph-J, Oph-K, Oph-L, Oph-M, and Oph-RingSW, respectively.
  Oph-M and the molecular ring Oph-RingSW are identified for the first time.
  Eight N$_2$H$^+$ peaks in clumps of L1688 are all co-located with dust peaks within one beam (60$''$).

  2) The observed C$_{2}$H to N$_{2}$H$^{+}$ ratio, [C$_{2}$H]/[N$_{2}$H$^{+}$], varies from 5 to 110, and is little affected by the uncertainties of derived temperatures.
  [C$_{2}$H]/[N$_{2}$H$^{+}$] in L1688 tends to be higher than that in L1689.
  For any given clump, [C$_{2}$H]/[N$_{2}$H$^{+}$] in the inner region tends to be lower than in the outer region.

  3) We modeled C$_{2}$H and N$_2$H$^+$ abundances with 1-D chemical models.
  The main difference between pure gas and gas-grain models is the dramatic drop in C$_2$H abundance in later stages of pure gas models due to destruction by reactions with CO, which will be depleted with the presence of dust.

  4) The diverging trends of abundances of C$_2$H and N$_2$H$^+$ in our gas-grain models make [C$_{2}$H]/[N$_{2}$H$^{+}$] a signpost of cloud conditions.
  For clouds with densities n$_H$ $\leq$ 10$^5$ cm$^{-3}$, [C$_{2}$H]/[N$_{2}$H$^{+}$] drops with time by $\rm \sim$2 order of magnitudes.
  In the gas with a density of n$_H$ $\geq$ 10$^5$ cm$^{-3}$ or higher, the [C$_{2}$H]/[N$_{2}$H$^{+}$] becomes more sensitive to density.

  5) The observed [C$_{2}$H]/[N$_{2}$H$^{+}$] values are the results of time evolution, accelerated at higher densities.
  For the relative low density regions in L1688 where only C$_2$H emission was detected, the gas should be chemically younger.

\acknowledgments

  This work is Supported by National Basic Research Program of China (973 program) No. 2012CB821800 and 2015CB857101,
  National Science Foundation of China No. 11373038,
  and CAS International Partnership Program, Grant No. 114A11KYSB20160008. 
  We thank the anonymous referee for his or her help. This paper was improved so much with the comments and suggestions. 
  The observation was made with the Delingha 13.7m telescope of the Qinghai Station of Purple Mountain Observatory$^{1}$, Chinese Academy of Sciences.
  We appreciate all the staff members of the observatory for their help during the observation.
  The telescope and the millimeter wave radio astronomy database is supported by the millimeter \& sub-millimeter wave laboratory of PMO.
  Di Li acknowledges the sponsorship provided by the Guizhou Normal University and support from the $''$CAS Interdisciplinary Innovation Team$''$.
  L.Q. is supported by the FAST FELLOWSHIP. 
  The FAST FELLOWSHIP is supported by Special Funding for Advanced Users, 
  budgeted and administrated by Center for Astronomical Mega-Science, Chinese Academy of Sciences (CAMS).
  We thank Ugo Hincelin for allowing us to use the Nautilus codes.
  We thank Dr. Juan Li for helping us with the C$_2$H hyperfine fittings.

\appendix

\section{The Critical Density Estimation Details}

  The critical density $n_{cr}$ is the density of molecular hydrogen
  at which the collisional de$-$excitation rate of the target molecules by hydrogen equals to the spontaneous de-excitation rate of the target molecule, i.e.,

\begin{equation}
     \textit{$n_{cr}C_{ul} = A_{ul}$}.
\end{equation}

  We define $\sigma$ as the collisional cross section,
  $v$ as the velocity of hydrogen molecules relative to target molecules,
  and $A$ as the Einstein $A$ coefficient (see next subsection).
  In addition, the relative velocity of hydrogen molecules can be estimated as

\begin{equation}
    \textit{$v = \sqrt{\frac{\rm k T}{m_{\rm H_2}}}\approx 2\rm\ km/s \left(\frac{T}{10\rm\ K}\right)^{1/2}$}.
\end{equation}

With these definitions, the critical density can be written as

\begin{equation}
    \textit{$n = 5\times 10^{10}A\left(\frac{\sigma}{10^{-16}\rm\ cm^{-2}}\right)^{-1}\left(\frac{\rm T}{10\rm\ K}\right)^{-1/2}\rm\ cm^{-3}$}.
\end{equation}

Values of the collisional cross section,
can be found in BASECOL\footnote{http://basecol.obspm.fr/index.php?page=pages/generalPages/home} (Dubernet et al.\ 2013),
which is part of the $''$Virtual Atomic and Molecular Data Center$''$ (VAMDC).
The spectroscopic data of BASECOL comes from
CDMS\footnote{http://basecol.obspm.fr/index.php?page=pages/generalPages/home} (the Cologne Database for Molecular Spectroscopy, M$\ddot{u\rm}$ller et al.\ 2005; M$\ddot{u\rm}$ller et al.\ 2001),
and JPL databases\footnote{http://spec.jpl.nasa.gov} (Pickett et al.\ 1998).
The original data of the C$_{2}$H and N$_2$H$^+$ collisional cross sections are from Spielfiedel et al.\ (2012) and Daniel et al.\ (2005), respectively.
Table \ref{critical_cch} and Table \ref{critical_n2hp} show the calculation results of critical densities $n_{cr}$ for C$_{2}$H N=1-0 and N$_2$H$^+$ J=1-0 hyperfine lines, respectively.
The critical density of N$_2$H$^+$ J=1-0 can also be found in the literature, for example, $1.5\times 10^{5}\rm \ $ cm$^{-3} $\textbf{ }\citep{Tanaka2013}.

\begin{table}
\centering
\begin{tiny}
  \caption{The C$_{2}$H N=1-0 critical densities. }
\vspace{1em}
  \begin{tabular}{lcccccc}
  \hline
  Transition                                       & Rest Frequency       &  $n_{10\rm K }$                     &  $n_{20\rm K}$                     &  $n_{30\rm K}$                    &   $n_{40\rm K}$  \\
                                                   &     (MHz)            &   (Unit)                     &     (Unit)                   &     (Unit)                  &     (Unit)    \\
  \hline
  C$_{2}$H N=1-0 J=$\frac{3}{2}$-$\frac{1}{2}$ F=1-1   & 87284.156            &     $4.4842 \times 10^{5}$  & 	$4.3544 \times 10^{5}$  &  	$4.9277 \times 10^{5}$  &	   $5.5602 \times 10^{5}$  \\
  C$_{2}$H N=1-0 J=$\frac{3}{2}$-$\frac{1}{2}$ F=2-1   & 87316.925            &     $1.7880 \times 10^{5}$  & 	$1.6694 \times 10^{5}$  &  	$1.8472 \times 10^{5}$  &	   $2.0518 \times 10^{5}$  \\
  C$_{2}$H N=1-0 J=$\frac{3}{2}$-$\frac{1}{2}$ F=1-0   & 87328.624            &     $2.8204 \times 10^{5}$  & 	$2.6018 \times 10^{5}$  &  	$2.8599 \times 10^{5}$  &	   $3.1623 \times 10^{5}$  \\
  C$_{2}$H N=1-0 J=$\frac{1}{2}$-$\frac{1}{2}$ F=1-1   & 87402.004            &     $2.1191 \times 10^{5}$  & 	$2.6415 \times 10^{5}$  &  	$3.1571 \times 10^{5}$  &	   $3.5970 \times 10^{5}$  \\
  C$_{2}$H N=1-0 J=$\frac{1}{2}$-$\frac{1}{2}$ F=0-1   & 87407.165            &     $1.4128 \times 10^{5}$  & 	$1.7610 \times 10^{5}$  &  	$2.1047 \times 10^{5}$  &	   $2.3980 \times 10^{5}$  \\
  C$_{2}$H N=1-0 J=$\frac{1}{2}$-$\frac{1}{2}$ F=1-0   & 87446.512            &     $4.2382 \times 10^{5}$  & 	$5.2829 \times 10^{5}$  &  	$6.3142 \times 10^{5}$  &	   $7.1942 \times 10^{5}$  \\
\hline
\label{critical_cch}
\end{tabular}
\end{tiny}
\end{table}

\begin{table}
\centering
\begin{tiny}
  \caption{The N$_2$H$^+$ J=1-0 critical densities. }
\vspace{1em}
  \begin{tabular}{lcccccc}
  \hline
  Transition                                       & Rest Frequency       &  $n_{10\rm K}$                     &  $n_{20\rm K}$                     &  $n_{30\rm K}$                    &   $n_{40\rm K}$  \\
                                                   &     (MHz)            &   (Unit)                     &     (Unit)                   &     (Unit)                  &     (Unit)    \\
  \hline
  N$_2$H$^+$ J=1-0 F$_{1}$=1-1 F=0-1                  & 93171.621          &     $3.3296 \times 10^{5}$  &   	$3.7340 \times 10^{5}$   &  	$3.9352 \times 10^{5}$   &	$4.0659 \times 10^{5}$  \\
  N$_2$H$^+$ J=1-0 F$_{1}$=1-1 F=2-1                  & 93171.917          &     $1.3319 \times 10^{6}$  &   	$1.4936 \times 10^{6}$   &  	$1.5741 \times 10^{6}$   &	$1.6264 \times 10^{6}$  \\
  N$_2$H$^+$ J=1-0 F$_{1}$=1-1 F=1-1                  & 93172.053          &     $1.3318 \times 10^{6}$  &   	$1.4935 \times 10^{6}$   &  	$1.5740 \times 10^{6}$   &	$1.6264 \times 10^{6}$  \\
  N$_2$H$^+$ J=1-0 F$_{1}$=2-1 F=2-1                  & 93173.480          &     $4.4392 \times 10^{5}$  &   	$4.9783 \times 10^{5}$   &  	$5.2468 \times 10^{5}$   &	$5.4210 \times 10^{5}$  \\
  N$_2$H$^+$ J=1-0 F$_{1}$=2-1 F=3-1                  & 93173.777          &     No Data  &     No Data   &    No Data   &  No Data  \\
  N$_2$H$^+$ J=1-0 F$_{1}$=2-1 F=1-1                  & 93173.967          &     $7.9897 \times 10^{5}$  &   	$8.9607 \times 10^{5}$   &  	$9.4439 \times 10^{5}$   &	$9.7576 \times 10^{5}$  \\
  N$_2$H$^+$ J=1-0 F$_{1}$=0-1 F=1-1                  & 93176.265          &     $9.9878 \times 10^{5}$  &   	$1.1201 \times 10^{6}$   &  	$1.1805 \times 10^{6}$   &	$1.2197 \times 10^{6}$  \\
\hline
\label{critical_n2hp}
\end{tabular}
\end{tiny}
\end{table}

\section{The Column Density Calculation Details}

  A simple solution to the radiative transfer equation is
\begin{equation}
   \textit{$T_{A} = [T_{x} - T_{bg}](1-e^{-\tau_{\nu}})$},
  \label{appendix_1}
\end{equation}
where the source temperature $T_{A}$, the excitation temperature $T_{x}$ and the background temperature $T_{bg}$ are all modified Planck functions.

  The relation between the optical depth $\tau_{\nu}$ at the frequency $\nu$ and the molecular column density in the upper level, $N_{u}$, is

\begin{equation}
  \textit{$\int \tau_{\nu} d\nu = \frac{A_{ul}c^{2}hN_{u}}{8 \pi kT_{x}\nu}$}.
  \label{appendix_2}
\end{equation}

  Here, $A_{ul}$ is the spontaneous radiative decay coefficient, $c$ is the speed of light, $h$ and $k$ are the Planck and Boltzmann constants, respectively.
  Combining Equation (\ref{appendix_1}) and Equation (\ref{appendix_2}),
  the $N_{u}^{0}$,
  which is the column density for the molecule in the upper level,
  can be calculated in the Rayleigh-Jeans approximation and optically thin limits as

\begin{equation}
  \textit{$N_{u}^{0} = \frac{8 \pi k \nu^{2}}{h c^{3} A_{ul}} \int T_{A} d\upsilon$}.
  \label{appendix_3}
\end{equation}

  The total column density can be obtained after some correction \citep{Li2002,Qian2012}
\begin{equation}
  \textit{$N_{\rm tot} = F_{u} F_{\tau} F_{b} N^0_{u}$},
\end{equation}
where $F_{u}$, $F_{\tau}$, and $F_b$ are the level correction factor, correction factor for opacity,
and the correction for the background.
$F_{u}$ can be written as
\begin{equation}
  \textit{$F_{u} = F_{RJ}\frac{Z}{(2J+1)}e^{\frac{hB_e J(J+1)}{kT_x}}$},
\end{equation}
where $Z$ is the partition function.
The correction factor $F_{RJ}$ is corrects for the difference between the Planck's Law and the Rayleigh-Jeans limit.
This difference is caused by the Taylor series expansion of $e^{\frac{h c}{\lambda k T_{x}}}$ in the low column density and low temperature assumptions.

  The correction factor $F_{RJ}$ is
\begin{equation}
  \textit{$F_{RJ} = \frac{1/(e^{h \nu/k T_{x}} -1)}{k T_{x}/h \nu} = \frac{h \nu}{k T_{x} (e^{h \nu/k T_{x}}-1)}$}.
  \label{appendix_9}
\end{equation}
  The opacity correction and background radiation correction are analysed.
  The correction factor accounting for opacity is \citep{Li2002}
\begin{equation}
  \textit{$F_{\tau} = \frac{\int \tau d\upsilon}{\int (1 - e^{-\tau}) d\upsilon}$},
  \label{appendix_10}
\end{equation}
  and the correction factor for background radiation is (Li 2002)
\begin{equation}
  \textit{$F_{b} = [1 - \frac{e^{h \nu/k T_{x}}-1}{e^{h \nu/k T_{bg}}-1}]^{-1}$}.
  \label{appendix_11}
\end{equation}
  For an optically thin spectrum, the $F_{\tau}$=1.
  For the correction for spectra with optical depth between 0.5 and 5, we use
\begin{equation}
  \textit{$F_{\tau} = \frac{\tau }{1 - e^{-\tau} }$}.
  \label{appendix_10_2}
\end{equation}

  Combining Equations (\ref{appendix_3}), (\ref{appendix_9}), (\ref{appendix_10}) (or \ref{appendix_10_2}), and (\ref{appendix_11}), the corrected column density of the molecule in upper level, $N_{u}$, is

%

\begin{equation}
  \textit{$N_{u} = N_{u}^{0} F_{RJ} F_{\tau} F_{b} =  \frac{8 \pi k \nu^{2}}{h c^{3} A_{ul}} \int T_{A} d\upsilon \frac{h \nu}{k T_{x} (e^{h \nu/k T_{x}}-1)} \frac{\tau}{1 - e^{-\tau}} [1 - \frac{e^{h \nu/k T_{x}}-1}{e^{h \nu/k T_{bg}}-1}]^{-1}$}.
  \label{appendix_12}
\end{equation}

  In our C$_{2}$H and N$_2$H$^+$ column density calculation, we assume that the C$_{2}$H and N$_2$H$^+$ spectra are all optically thin, the gas is in LTE, and $T_{bg} = 2.73\ \rm K$.
  Temperatures from the IRAS temperature map in COMPLETE are used as $T_{x}$, which is the excitation temperature.
  $\int T_{A} d\upsilon$ is the integrated intensity.

  Table \ref{abundance_para} shows the parameters in Equation \ref{appendix_12}.
  The $A_{ul}$ for C$_{2}$H comes from Tucker et al.\ (1974) and the $A_{ul}$ for N$_2$H$^+$ comes from Turner (1974).
  $B_{e}$ values come from JPL Molecular Spectroscopy (Pickett et al.\ 1998).


\begin{table}
\centering
  \caption{Parameters used for calculating column density}
\vspace{1em}
  \begin{tabular}{lccccc}
  \hline
  Molecular  &  Frequency               & $A_{ul}$                &   $B_{e}$   \\
             &    (GHz)                 &  (s$^{-1}$)             & (MHz) \\
  \hline
  C$_{2}$H       &   87.3486                &  $1.66 \times 10^{-6}$  &  43674.534           \\
  N$_2$H$^+$       &   93.1738                &  $3.84 \times 10^{-5}$   &  46586.867   \\
  \hline
  Constant   &  k (J/K)                 &  h  (J.S)               &   c (m/s)                \\
             & $1.38 \times 10^{-23}$   &  $6.63 \times 10^{-34}$ &  $3 \times 10^{8}$       \\
  \hline
  \label{abundance_para}
\end{tabular}
\end{table}

  The total column density $N_{tot}$ can be calculated by

\begin{equation}
  \textit{$N_{tot} = N_{u} \frac{Z}{2J+1} e^{\frac{h B_{e} J (J+1)}{k T_{x}}}$},
  \label{appendix_14}
\end{equation}

  where

\begin{equation}
  \textit{$Z = \sum_ 0 ^\infty (2J+1) e^{- \frac{h B_{e}J(J+1) }{kT_{ex}} }$}.
  \label{appendix_15}
\end{equation}

Here, as an approximation, we use

\begin{equation}
  \textit{$Z' = \sum_ 0 ^{100000} (2J+1) e^{- \frac{h B_{e}J(J+1) }{kT_{ex}} }$},
  \label{appendix_z}
\end{equation}

  where the errors will be less than 1\%.

  These equations are for N$_2$H$^+$.
  Though the transition of C$_{2}$H is N=1-0, where N is the total angular momentum exclusive of spin,
  the same calculation is also suitable for C$_{2}$H.
  One similar calculation was presented in Tucker et al.\ (1974) for the first detection of C$_{2}$H in a number of Galactic sources.

\section{The interpolation of including our observed [C$_{2}$H]/[N$_{2}$H$^{+}$] values in the modeled data}

  The chemical model was run with four temperatures (T$_1$=10 K, T$_2$=20 K, T$_3$=30 K, and T$_4$=40 K) and
  four volume densities (n$_1$=5.0$\times$10$^3$ cm$^{-3}$, n$_2$=5.0$\times$10$^3$ cm$^{-3}$, n$_3$=5.0$\times$10$^3$ cm$^{-3}$, and n$_4$=5.0$\times$10$^3$ cm$^{-3}$).
  For one temperature and one volume density,
  we obtained a set of [C$_{2}$H]/[N$_{2}$H$^{+}$] values
  (namely [C$_{2}$H]/[N$_{2}$H$^{+}$]$_{x,y,z}$, x=1, 2, 3, or 4 is the index of the temperature, y=1, 2, 3, or 4 is the index of the volume density, z is the index in the set) and
  corresponding chemical ages (namely t$_{x,y,z}$, x, y, and z have the same meanings as those in [C$_{2}$H]/[N$_{2}$H$^{+}$]$_{x,y,z}$) from the chemical model.
  With the four temperatures and four volume densities, we obtained 16 sets of [C$_{2}$H]/[N$_{2}$H$^{+}$] values and corresponding chemical ages.
  For \textbf{\sout{a} an} observed [C$_{2}$H]/[N$_{2}$H$^{+}$] value, namely [C$_{2}$H]/[N$_{2}$H$^{+}$]$_{oph}$, under the given temperature and volume density, its chemical age can be calculated by linear interpolation as:

\begin{equation}
  \textit{ \rm t$_{oph,x,y}$ = t$_{x,y,z}$ + (t$_{x,y,z+1}$ - t$_{x,y,z}$) $\times$ K$_1$},
  \label{appendix_inter_1}
\end{equation}
  where K$_1$ is
\begin{equation}
  \textit{\rm K$_1$ = ([C$_2$H]/[N$_2$H$^+$]$_{oph}$ - [C$_2$H]/[N$_2$H$^+$]$_{x,y,z}$ )/( [C$_2$H]/[N$_2$H$^+$]$_{x,y,z+1}$ - [C$_2$H]/[N$_2$H$^+$]$_{x,y,z}$) }.
  \label{appendix_inter_2}
\end{equation}
  Here, t$_{oph,x,y}$ is the obtained chemical age,
  [C$_2$H]/[N$_2$H$^+$]$_{x,y,z}$ and [C$_2$H]/[N$_2$H$^+$]$_{x,y,z+1}$ are the abundance values which are the two values closest to the observed [C$_2$H]/[N$_2$H$^+$]$_{oph}$ and
\begin{equation}
  \textit{\rm [C$_2$H]/[N$_2$H$^+$]$_{x,y,z}$ $<$ [C$_2$H]/[N$_2$H$^+$]$_{oph}$, [C$_2$H]/[N$_2$H$^+$]$_{x,y,z+1}$ $>$ [C$_2$H]/[N$_2$H$^+$]$_{oph}$},
  \label{appendix_inter_3}
\end{equation}
  t$_{x,y,z}$ and t$_{x,y,z+1}$ are the chemical ages corresponding to [C$_2$H]/[N$_2$H$^+$]$_{x,y,z}$ and [C$_2$H]/[N$_2$H$^+$]$_{x,y,z+1}$, respectively.
  For the 16 sets, we have 16 values of t$_{oph,x,y}$ in which x=1, 2, 3, or 4 and y=1, 2, 3, or 4.
  If the IRAS dust temperature of [C$_2$H]/[N$_2$H$^+$]$_{oph}$ is T$_{oph}$, the estimated chemical age of [C$_2$H]/[N$_2$H$^+$]$_{oph}$ under one volume density is
\begin{equation}
  \textit{ \rm t$_{oph}$ = t$_{oph,x,y}$ + (t$_{oph,x,y+1}$ - t$_{oph,x,y}$) $\times$ K$_2$},
  \label{appendix_inter_4}
\end{equation}
  where K$_2$ is
\begin{equation}
  \textit{\rm K$_2$ = ( T$_{oph}$ - T$_{x,y,z}$ )/( T$_{x,y,z+1}$ - T$_{x,y,z}$ ) }.
  \label{appendix_inter_5}
\end{equation}
  Here, t$_{oph}$ is the determined chemical age under the given volume density,
  T$_{x,y,z}$ and T$_{x,y,z+1}$ are two temperatures which are closest to T$_{oph}$ and
\begin{equation}
  \textit{\rm T$_{x,y,z}$ $<$ T$_{oph}$, T$_{x,y,z+1}$ $>$ T$_{oph}$},
  \label{appendix_inter_6}
\end{equation}
  t$_{oph,x,y}$ and t$_{oph,x,y+1}$ are two chemical ages which come from Equation \ref{appendix_inter_1} under one same volume density and temperatures of T$_{x,y,z}$ and T$_{x,y,z+1}$, respectively.


\begin{thebibliography}{}
\bibitem[\protect\citeauthoryear{{Andr\'{e}} et al.}{2007}]{Andre2007} {Andr\'{e} Ph}., Belloche A., Motte F., \& Peretto N. 2007, A\&A, 472, 519-535
\bibitem[\protect\citeauthoryear{Daniel et al.}{2005}]{Daniel2005} Daniel, F., Dubernet, M.-L., Meuwly, M., Cernicharo, J., \& Pagani, L. 2005, MNRAS, 363, 1083-1091
\bibitem[\protect\citeauthoryear{Di Francesco et al.}{2004}]{DiFrancesco2004} Di Francesco J., Andr\`{e} P., Myers P., 2004, ApJ, 617, 425-438
\bibitem[\protect\citeauthoryear{Dubernet et al.}{2013}]{Dubernet2013} Dubernet, M.-L., Alexander, M. H., Ba, Y. A. et al. 2013, A\&A, 553, A50
\bibitem[\protect\citeauthoryear{Hersant et al.}{2009}]{Hersant2009} Hersant, F., Wakelam, V., Dutrey, A., Guilloteau, S. \& Herbst, E. 2009, A\&A, 493, 49H
\bibitem[\protect\citeauthoryear{Hincelin et al.}{2011}]{Hincelin2011} Hincelin, U., Wakelam, V., Hersant, F., et al. 2011, A\&A, 530, A61
\bibitem[\protect\citeauthoryear{Hincelin et al.}{2013}]{Hincelin2013} Hincelin, U., Wakelam, V., Commercon, B., Hersant, F. \& Guilloteau, S. 2013, ApJ, 775, 44
\bibitem[\protect\citeauthoryear{Johnstone et al.}{2004}]{Johnstone2004} Johnstone, D., Di Francesco, J., \& Kirk, H. 2004, ApJ, 611, L45-L48
\bibitem[\protect\citeauthoryear{Keto \& Rybicki}{2010}]{Keto2010} Keto E., \& Rybicki G. 2010, ApJ, 716, 1315-1322
\bibitem[\protect\citeauthoryear{Kristensen et al.}{2013}]{Kristensen2013} Kristensen, L. E., Klaassen, P. D., Mottram, J. C., Schmalzl, M., \& Hogerheijde, M. R. 2013, A\&A, 549, L6
\bibitem[\protect\citeauthoryear{Lada \& Lada}{2003}]{Lada2003} Lada C. J., \& Lada E. A. 2003, ARA\&A, 41, 57L
\bibitem[\protect\citeauthoryear{Li}{2002}]{Li2002} Li, D. 2002, Ph.D. Thesis, cornell University
\bibitem[\protect\citeauthoryear{Li et al.}{2012}]{Li2012} Li, J., Wang, J., Gu, Q., Zhang, Z., \& Zheng X 2012, ApJ, 745, 47
\bibitem[\protect\citeauthoryear{Liseau et al.}{1999}]{Liseau1999} Liseau R., White G. J., Larsson B. et al. 1999, A\&A, 344, 342-354
\bibitem[\protect\citeauthoryear{Loinard et al.}{2008}]{Loinard2008} Loinard, L., Torres, R. M., Mioduszewski, A. J., \& Rodr$\acute{i}$guez L. F. 2008, ApJ, 675, L29-L32
\bibitem[\protect\citeauthoryear{Loren et al.}{1990}]{Loren1990} Loren, R. B., Wootten, A., \& Wilking, B. A. 1990, ApJ, 365, 269-286
\bibitem[\protect\citeauthoryear{Motte et al.}{1998}]{Motte1998} Motte, F., Andr\`{e}, P., \& Neri, R. 1998, A\&A, 336, 150-172
\bibitem[{{M{\"u}ller} et al.}(2005)]{Muller2005} M{\"u}ller, H. S. P., {Schl{\"o}der}, F., Stutzki, J., \& Winnewisser, G. 2005, JMoSt, 742, 215M
\bibitem[{{M{\"u}ller} et al.}(2001)]{Muller2001} M{\"u}ller, H. S. P., Thorwirth, S., Roth, D. A., \& Winnewisser, G. 2001, A\&A, 370, L49-L52
\bibitem[\protect\citeauthoryear{Nutter et al.}{2006}]{Nutter2006} Nutter, D., Ward-Thompson, D., \& Andr\`{e}, P. 2006, MNRAS, 368, 1833N
\bibitem[\protect\citeauthoryear{Padgett et al.}{2008}]{Padgett2008} Padgett, D. L., Rebull, L. M., Stapelfeldt, K. R. et al. 2008, ApJ, 672, 1013-1037
\bibitem[\protect\citeauthoryear{Pattle et al.}{2015}]{Pattle2015} Pattle, K., Ward-Thompson, D., Kirk, J. M. et al. 2015, MNRAS, 450, 1094-1122
\bibitem[\protect\citeauthoryear{Pickett et al.}{1998}]{Pickett1998} Pickett, H. M., Poynter, R. L., Cohen, E. A. et al. 1998, JQSRT, 60, 883P
\bibitem[\protect\citeauthoryear{Qian et al.}{2012}]{Qian2012} Qian, L., Li D., \& Goldsmith P. 2012, ApJ, 760, 147
\bibitem[\protect\citeauthoryear{Ridge et al.}{2006}]{Ridge2006} Ridge, N. A., Di Francesco, J., Kirk, H. et al. 2006, AJ, 131, 2921-2933
\bibitem[\protect\citeauthoryear{Semenov et al.}{2010}]{Semenov2010} Semenov, D., Hersant, F., Wakelam, V. et al. 2010, A\&A, 522, A42
\bibitem[\protect\citeauthoryear{Shan et al.}{2012}]{Shan2012} Shan, W., Yang, J., Shi, S. et al. 2012, IEEE, vol.2, issue 6, 593-604
\bibitem[\protect\citeauthoryear{Schnee et al.}{2005}]{Schnee2005} Schnee, S. L., Ridge, N. A., Goodman, A. A., \& Li, J. G. 2005, ApJ, 634, 442-450
\bibitem[\protect\citeauthoryear{Shirley et al.}{2015}]{Shirley2015} Shirley, Y. L. 2015, PASP, 127, 299S
\bibitem[\protect\citeauthoryear{Spielfiedel et al.}{2012}]{Spielfiedel2012} Spielfiedel, A., Feautrier, N., Najar, F. et al. 2012, MNRAS, 421, 1891-1896
\bibitem[\protect\citeauthoryear{Stanke et al.}{2006}]{Stanke2006} Stanke, T., Smith, M. D., Gredel, R., \& Khanzadyan, T. 2006, A\&A, 447, 609-622
\bibitem[\protect\citeauthoryear{Tanaka et al.}{2013}]{Tanaka2013} Tanaka, T., Nakamura F., Awazu F. et al. 2013, ApJ, 778, 34
\bibitem[\protect\citeauthoryear{Tucker et al.}{1974}]{Tucker1974} Tucker, K. D., Kutner, M. L., \& Thaddeus, P. 1974, ApJ, 193, L115-L119
\bibitem[\protect\citeauthoryear{Turner}{1974}]{Turner1974} Turner, B. E. 1974, ApJ, 193, L83-L87
\bibitem[\protect\citeauthoryear{Turner}{1995}]{Turner1995} Turner, B. E. 1995, ApJ, 449, 635T
\bibitem[\protect\citeauthoryear{Vasyunina}{2012}]{Vasyunina2012} Vasyunina, T., Vasyunin, A. I., Herbst, Eric, \& Linz H. 2012, ApJ, 751, 105V
\bibitem[\protect\citeauthoryear{Wakelam et al.}{2012}]{Wakelam2012} Wakelam, V., Herbst, E., Loison, J.-C. et al. 2012, ApJS, 199, 21w
\bibitem[\protect\citeauthoryear{Young et al.}{2006}]{Young2006} Young, K. E., Enoch, M. L., Evans II, N. J. et al. 2006, ApJ, 644, 326-343
\bibitem[\protect\citeauthoryear{Zuo et al.}{2011}]{Zuo2011} Zuo, Y., Li, Y., Sun, J. et al. 2011, RAA, 35, 439-446
\end{thebibliography}
\end{document}